\documentstyle[preprint,pre,aps]{revtex}

\tightenlines

\begin{document}

{
	\draft
	\title{Associative memory storing an extensive number of patterns based on a network of oscillators with distributed natural frequencies in the presence of external white noise}
	\author{Masahiko Yoshioka\thanks{Electronic address: myosioka@brain.riken.go.jp} and Masatoshi Shiino\thanks{Electronic address: mshiino@ap.titech.ac.jp}}
	\date{{March 19, 1999}}
	\address{Department of Applied Physics Tokyo Institute of Technology Ohokayama, Meguro-ku, Tokyo 152, Japan}
	\maketitle
}

\begin{abstract}
We study associative memory based on temporal coding in which successful retrieval is realized as an entrainment in a network of simple phase oscillators with distributed natural frequencies under the influence of white noise.
The memory patterns are assumed to be given by uniformly distributed random numbers on $[0,2\pi )$ so that the patterns encode the phase differences of the oscillators.
To derive the macroscopic order parameter equations for the network with an extensive number of stored patterns, we introduce the effective transfer function by assuming the fixed-point equation of the form of the TAP equation, which describes the time-averaged output as a function of the effective time-averaged local field.
Properties of the networks associated  with synchronization phenomena for a discrete symmetric natural frequency distribution with three frequency components are studied based on the order parameter equations, and are shown to be in good agreement with the results of numerical simulations.
Two types of retrieval states are found to occur with respect to the degree of synchronization, when the size of the width of the natural frequency distribution is changed.
\end{abstract}

{}

\section{Introduction}

Recently a population of neurons in the cat visual cortex has been reported to exhibit synchronized firings in a stimulus dependent manner\cite{gray,eckhorn}.
The occurrence of correlations in firing times of neurons seems to be a ubiquitous phenomenon in real nervous systems.
The role of such synchronized firings for information processing in the brain has been attracting growing interest of researchers, and several authors have suggested neural network models based on the concept of temporal coding, where information of a neuron is represented by its firing times.
Indeed, to explain the experimental findings that visual information of an external object is processed with being divided into several pieces of information, several authors\cite{malsburg,malsburg2,milner,damasio} have suggested that the synchronized firings of neurons may serve as a linker for those pieces of information.

The problem of investigating how an associative memory is realized in real nervous systems as well as of constructing biologically relevant models is of central concern of neuroscientists.
Since the establishment of  systematic theories of associative memory for networks with an energy function that is ensured by assuming symmetric synaptic couplings\cite{hopfield,amit,shiino,kuhn,kuhn2,gardner,kuhn3,coolen,fukai,waugh}, several attempts have been made to make models as biologically plausible as possible\cite{cugliandolo,cugliandolo2,griniasty}.
Previously we investigated the effects of asymmetric couplings\cite{peretto} for memorizing pre-synaptic and post-synaptic activities, which are incorporated into the standard symmetric Hebb learning rule\cite{hebb}, by studying networks of analog neurons\cite{yoshioka,yoshioka2}, whose continuous-time dynamics involves a positive-valued transfer function representing the mean firing rates of a neuron as a function of membrane potential.

Instead of working with the concept of rate coding based on the idea that neuronal information is represented by mean firing rate of a single neuron, one may be concerned with the concept of temporal coding, when considering that spatio-temporal patterns of neuronal firings will make information carried by a population of neurons much richer than spatial patterns alone.
A spiking neuron is considered to be one of the candidates for implementation of temporal coding\cite{mass,hansel,vreeswijk,gerstner,gerstner2,herz,treves,bressloff1,bressloff2,carson}.
The time evolution of membrane potential that is generated in response to an injected synaptic electric current of a spiking neuron is described by such nonlinear dynamics as Hodgkin-Huxley equation\cite{hodgkin}, FitzHugh-Nagumo equation\cite{fitzhugh,nagumo}, or the equation of an integrated-and-fire neuron.
Spiking neurons in a network are generally supposed to interact with each other via pulses generated in the firing events occurring in the pre-synaptic neuron.
We have shown previously that even in the presence of time delays in transmission of the pulses an associative memory based on a network of spiking neurons can be realized  by assuming a simple Hebb-type learning rule alone, and that the memory retrieval accompanies synchronized firings of neurons.
The dynamics of such associative memory was analyzed by means of sublattice method in our previous paper\cite{yoshioka3}.

In the previous analysis\cite{yoshioka3} we assumed that every neuron shares identical characteristics to exhibit the same reaction in response to the same injected current.
In real nervous systems, however, neurons in a network may have their own individual characteristics.

The problem of whether temporal coding functions robustly in the presence of certain heterogeneties of neurons will be of particular interest.
For the purpose of investigating such a problem for the phenomenon of synchronized oscillations in associative memory neural networks, we consider it appropriate to deal with simple  models of coupled oscillators with distributed natural frequencies and external noise.
It is well known that, under certain conditions, a population of oscillators with distributed natural frequencies is allowed to get partially entrained in such a way that oscillators with a natural frequency near the central frequency become to oscillate synchronously at identical frequency as a result of cooperative interactions\cite{kuramoto,sakaguchi,daido}.

Kuramoto\cite{kuramoto} showed that the dynamics of this kind of network of oscillators with sufficiently week interactions can be  reduced to a simple phase dynamics.
Supposing that neurons in a network are treated as phase oscillators, associative memory has been shown to be realized under a simple learning rule of the Hebb-type either in the case of a finite number of stored patterns\cite{arenas,sakaguchi} or in the case with a single natural frequency\cite{cook,aoyagi}.
Satisfactory analysis of the case with a distribution of natural frequency and extensively many stored patterns has been far less conducted.
Quite recently we have reported the study of a deterministic phase oscillator network with a distribution of natural frequencies where an extensive number of binary patterns ($\pm 1$) are stored with use of the Hebb learning rule\cite{yamana}.

The main purpose of the present study is the theoretical analysis of associative memory based on temporal coding with use of networks of phase oscillators in the more general case where the number of stored patterns that are given by uniformly distributed random numbers on $[0,2\pi )$ is extensive, natural frequencies of the oscillators are distributed according to a certain distribution function, and furthermore external white noise is added to the system.

While one can analyze a phase oscillator network with a single natural frequency by means of the replica method that makes full use of its associated energy function, one cannot resort any more to the standard method of  statistical physics based on the existence of an energy function in the case of networks with  a distribution of natural frequencies.

One can, however, use the Self-Consistent Signal-to-Noise Analysis (SCSNA)\cite{shiino2,shiino3,yoshioka,yoshioka2,yoshioka3} to deal with general cases without energy functions.
To apply the SCSNA it is necessary to know fixed-point equations describing the equilibrium states of the network.
When considering such equations in stochastic systems, we may take advantage of the concept of the TAP equation\cite{thouless,mezard}.

The TAP equation is known to exist for the SK model of spin glasses\cite{mezard,kirkpatrick,plefka} and the Hopfield model of an Ising spin neural networks\cite{amit,mezard,nakanishi}, and to represent a functional relationship between the thermal or time average of each spin in equilibrium and its corresponding effective local field that involves the so called Onsager reaction term\cite{mezard}.
Usefulness of the TAP equation in deriving the order parameter equations of associative memory networks is attributed to the fact that the resulting equations of the replica calculations by AGS is recovered by the result of application of the SCSNA to the TAP equation\cite{shiino3}, where the Onsager reaction term is canceled exactly by the appearance of the renormalized output term of the SCSNA\cite{shiino3}.
We note that the TAP equation of the naive mean field model with the interactions given by the Hebb learning rule defines to an analog network equation with the transfer function $\tanh{\left ( {\beta h} \right )}$\cite{shiino3}.

We first evaluate an analogue of the TAP equation of the naive mean field type for our model with a distribution of natural frequencies by dealing with the Fokker-Plank equation.
In order to obtain the Onsager reaction term we compute the free energy of the network without a distribution of natural frequencies to derive the TAP equation by following the method of Plefka\cite{plefka} and Nakanishi\cite{nakanishi}.
Then we assume that the TAP-like equation also exists with the Onsager reaction term remaining the same even for networks with  a distribution of natural frequencies, and that such TAP-like equation defines an effective transfer function to which the SCSNA is applicable to obtain the order parameter equations.

The present paper is organized as follows.
In {{{section}~\ref{model}}} we introduce a neural network of simple phase oscillators and describe how the network functions as an associative memory based on  a simple learning rule of the Hebb-type.
In {{{section}~\ref{derivation}}} we give a theoretical analysis to derive the macroscopic order parameter equations describing the long time behavior of the system.
On the basis of the order parameter equations, in {{{section}~\ref{application}}} we investigate properties of memory retrieval accompanying synchronization in the networks by assuming a discrete symmetric natural frequency distribution with three frequency components.
Results of numerical simulations are presented showing good agreement with those of theoretical analysis.
In {{{section}~\ref{discussion}}}, comparing our work with those of other researchers conducted previously, we summarize the results of the present study.

\section{Neural networks of phase oscillators with distributed natural frequencies}{\label{model}}

The system under consideration is a network of $N$ phase oscillators subjected to external white noise, whose dynamics is expressed as
\begin{equation}
\dot{\phi_i}=\omega_i-\sum_{j\neq i}^N J_{ij}\sin{\left ( {\phi_i-\phi_j-\beta_{ij}} \right )}+\eta_i(t),{\label{system}}
\end{equation}
where $\phi_i$ and $\omega_i$ represent the phase and the natural frequency of oscillator $i$ respectively.
$\beta_{ij}$ and $J_{ij}$ represent a certain phase shift and the strength of coupling between oscillator $i$ and $j$ respectively.
The Gaussian white noise $\eta_i(t)$ is assumed to satisfy ${\big \langle}\eta_i(t){\big \rangle}=0$ and ${\big \langle}\eta_i(t)\eta_j(t'){\big \rangle}=2D\delta_{ij}\delta{\left ( {t-t'} \right )}$.

When {$\omega_i=0\ (i=1,\ldots,N)$}, $c_{ij}=J_{ij}\exp{\left ( {i\beta_{ij}} \right )}$ satisfy $c_{ij}={{{c_{ji}}}^\ast}$ with $\ast$ denoting complex conjugation and $J_{ij}$ take real values, the system~{{(\ref{system})}} has the energy function:
\begin{equation}
H{\left ( {{\left\{ {\phi_i} \right \}}} \right )}=-\frac{1}{2}\sum_{i\neq j}J_{ij}\cos{\left ( {\phi_i-\phi_j-\beta_{ij}} \right )}.{\label{hamiltonian}}
\end{equation}
Then one has an equilibrium probability distribution $\rho{\left ( {{\left\{ {\phi_i} \right \}}} \right )}$ proportional to $\exp{\left ( {-H{\left ( {{\left\{ {\phi_i} \right \}}} \right )}/D} \right )}$.
In the case of $D=0$, the function~{{(\ref{hamiltonian})}} becomes a Lyapunov function of the system and hence the state of the system eventually settles into a certain fixed point attractor after a long time.

In the present study we assume natural frequencies to be distributed accordingly to an even distribution function $g(\omega)=g(-\omega)$ so that the average of natural frequencies become zero without loss of generality.
To store $P$ quenched random patterns $\theta_i^\mu\ (i=1,\ldots,N,\ \mu=1,\ldots,P)$ chosen from the uniform distribution on the interval $[0,2\pi)$, we assume the Hebb type learning rule, and set the parameter $\beta_{ij}$ and real valued $J_{ij}$ such that 
\begin{equation}
c_{ij}=J_{ij}\exp{\left ( {i\beta_{ij}} \right )}=\left\{\begin{array}{lc}\frac{1}{N}\sum_{\mu=1}^P\xi_i^\mu{{{\xi_j^\mu}}^\ast}&i\neq j\\0&i=j\end{array}\right. ,{\label{hebb}}
\end{equation}
where $\xi_i^\mu=\exp{\left ( {i\theta_i^\mu} \right )}$.
This definition of couplings gives networks the following properties:
\begin{itemize}
\item In successful retrieval an entrainment occurs, where synchronized oscillators satisfy $\phi_i-\phi_j\approx\theta_i^\mu-\theta_j^\mu$ with a target pattern~$\mu$ recalled.
(Note that if $\phi_i$ is the solution of dynamics~{{(\ref{system})}}, uniformly shifted phase $\tilde{\phi_i}=\phi_i+c$ also becomes its solution.
What matters is not the phase itself but their difference $\phi_i-\phi_j$.)
\item In the case of unsuccessful retrieval, all the oscillators fail to synchronize, running at their own natural frequencies.
\end{itemize}

To measure the distance between the pattern~$\mu$ and the state of the system, we introduce the overlap for pattern~$\mu$:
\begin{equation}
m^\mu(t)=\frac{1}{N}\sum_i{{{\xi_i^\mu}}^\ast} z_i(t),{\label{overlap}}
\end{equation}
where we denote $e^{i\phi(t)}$ by $z_i(t)$.
Then by use of the local field:
\begin{equation}
h_i(t)=\sum_{j\ne i} c_{ij}z_j(t)=\sum_\mu \xi_i^\mu m^\mu(t)-\alpha z_i,
\end{equation}
the dynamics~{{(\ref{system})}} is rewritten as
\begin{equation}
\dot{\phi_i}=\omega_i-{\mbox{Re}}{\left\{ {h_i(t)} \right \}}\sin\phi_i+{\mbox{Im}}{\left\{ {h_i(t)} \right \}}\cos\phi_i+\eta_i(t),{\label{nativedynamics}}
\end{equation}
where $\alpha$ denotes the loading rate $P/N$.
From {{{Eq.}~{(\ref{nativedynamics})}}} it is easy to see that the learning rule~{{(\ref{hebb})}} indeed realizes the above mentioned properties if the number of stored patterns is finite($\alpha=0$) and {$\omega_i=0\ (i=1,\ldots,N)$} .
The memory retrieval accompanying synchronization can also occur for $\alpha>0$ even in the presence of a distribution of natural frequencies.

\section{Macroscopic order parameter equations}{\label{derivation}}

Behaviors of associative memory networks depend crucially on the nature of the local fields or the neurons of oscillators, because the updating rule for the time evolution of the system is based on the local fields as is seen in {{{Eq.}~{(\ref{nativedynamics})}}}.
When the long time behavior of a network is described by fixed point type attractors, the relation between the resulting output state of a neuron and the corresponding local fields becomes essential for determining equilibrium properties of the networks.
Such a relation is naturally introduced in the case of deterministic analog networks where neurons are characterized by transfer functions describing the input-output relation.

For such stochastic systems~{{(\ref{system})}} as Ising spin networks  equilibrium fixed-point equations called the TAP equations are known to exist as expressing the relation between time average of each output of a neuron and its corresponding effective local fields involving the so called Onsager reaction term that is proportional to the time-averaged output.
Once the TAP-like equation level description is available, the SCSNA, in which one compute the variance of the cross-talk noises in the local field as a result of storing an extensive number of patterns, can be applied to obtain the order parameter equations in the limit $N\rightarrow\infty$.

\subsection{Effective transfer function based on time-averaged local field and the TAP-like equation}

If {$\omega_i=0\ (i=1,\ldots,N)$}\ and $D=0$, the state of the network eventually settles into an equilibrium state given by a fixed point attractor owing to the existence of the Lyapunov function~{{(\ref{hamiltonian})}}, and then the local fields do not fluctuate in time.
Even in the presence of external white noise ($D\neq 0$), the local fields also get fixed in time after a long time, provided $\alpha=P/N=0$.
When the local fields are fixed over the time change due to the law of large numbers, theoretical treatment becomes simple because one can reduce the many body problem to a single body problem.

\begin{figure}
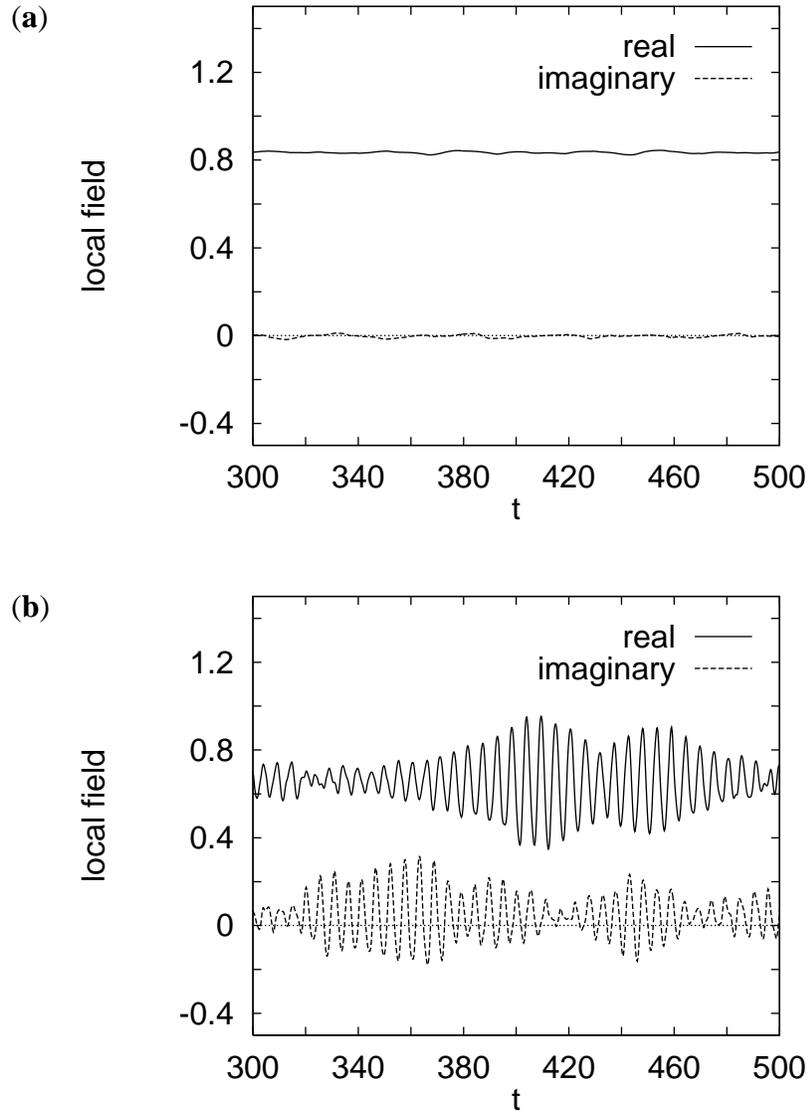

{}{}
{}{}
{}
\caption{The typical time evolution of the local fields observed in numerical simulations with $N=8000,\alpha=0.02,D=0$.
Natural frequencies are chosen so as to obey {{\bf (a)}}\ a Gaussian distribution $g(\omega)=\exp{\left ( {-\omega^2/2\sigma^2} \right )}/\sqrt{2\pi\sigma^2}$ with $\sigma=0.3$  and {{\bf (b)}}\ $g(\omega)=0.15\delta{\left ( {\omega+1.4} \right )}+0.7\delta{\left ( {\omega} \right )}+0.15\delta{\left ( {\omega-1.4} \right )}$.}
{\label{fluctuation}}
\end{figure}

In more general case where $\alpha>0$ and/or natural frequencies are distributed,  the local fields may fluctuate even with a large number of oscillators as can be  shown in the numerical simulation illustrated in {{{Fig.}~\ref{fluctuation}}}.
The fluctuations seem to be aperiodic and rigorous analysis of such fluctuations is quite difficult.
To deal this situation we are forced to resort to a certain approximation by confining ourselves only to the near equilibrium behavior of the system: {\it we replace the time-dependent local fields by their time-averaged ones:}
\begin{equation}
h_i(t)\approx\overline{h_i}=\sum_{j\ne i} c_{ij} \overline{z_j},
\end{equation}
where the over bar represents the time average at near equilibrium.

Once we apply this approximation, we are allowed to treat each interconnected oscillator as an element obeying the dynamics of one degree:
\begin{equation}
\dot{\phi_i}=\omega_i-{\mbox{Re}}{\left ( {\overline{h_i}} \right )}\sin\phi_i+{\mbox{Im}}{\left ( {\overline{h_i}} \right )}\cos\phi_i+\eta_i(t),{\label{replace}}
\end{equation}
which is expected to describe the behavior of the oscillator $i$ near equilibrium.
It turns out that the time average $\overline{z_i}$ at equilibrium can be expressed as a function of $\omega_i$ and $\overline{h_i}$:  $\overline{z_i}={f{\left ( {{\omega_i,\overline{h_i},\overline{h_i}^\ast}} \right )}}$.
We call that the transfer function in the case of network of analog neurons.
It is easy to show that the transfer function ${f{\left ( {{\omega,\overline{h},\overline{h}^\ast}} \right )}}$ satisfies
\begin{equation}
{f{\left ( {{\omega,r e^{i\theta},{\left ( {re^{i\theta}} \right )}^\ast}} \right )}}=e^{i\theta}{f{\left ( {{\omega,r,r}} \right )}}.{\label{circulate}}
\end{equation}
Hence it suffices to calculate ${f{\left ( {{\omega,r,r}} \right )}}$ to obtain ${f{\left ( {{\omega,r e^{i\theta},{\left ( {r e^{i\theta}} \right )}^\ast}} \right )}}$.
In the absence of white noise ($D=0$), we can easily obtain the transfer function for real-valued local field $r$\cite{kuramoto}: 
\begin{equation}
f(\omega,r,r)=\left \{
	\begin{array}{lc}
	\frac{i}{r}{\left ( {\omega+\sqrt{\omega^2-{r}^2}} \right )}&\omega<-r\\
	\frac{1}{r}{\left ( {i\omega+\sqrt{{r}^2-\omega^2}} \right )}&-r<\omega<r\\
	\frac{i}{r}{\left ( {\omega-\sqrt{\omega^2-{r}^2}} \right )}&r<\omega
	\end{array}
\right . .{\label{nod}}
\end{equation}
In deriving {{{Eq.}~{(\ref{nod})}}} in the case of $-r<\omega<r$ we used $\dot{\phi}=0$ together with $d \dot{\phi}/d \phi <0$.
In the case of $\omega<-r$ or $r<\omega$ we computed the time average of $\exp{\left ( {i\phi(t)} \right )}$ over the period $T_0$ of the periodic oscillations of $\phi$:
\begin{equation}
{f{\left ( {{\omega,r,r}} \right )}}=\frac{1}{T_0}\int_0^{T_0}\exp{\left ( {i\phi(t)} \right )} dt=\frac{\int_0^{2\pi}\frac{e^{i\phi}}{\dot{\phi}} d\phi}{\int_0^{2\pi}\frac{1}{\dot{\phi}} d\phi}=\frac{\int_0^{2\pi}\frac{e^{i\phi}}{\omega-r\sin \phi}d\phi}{\int_0^{2\pi}\frac{1}{\omega-r\sin\phi}d\phi}.
\end{equation}

In {{{Fig.}~\ref{field}}} we illustrate the shape of the transfer function ${f{\left ( {{\omega,{\overline{h}},{\overline{h}}^\ast}} \right )}}$ that is obtained from {{{Eqs.}~{(\ref{circulate})}}} and {{(\ref{nod})}}.

In the presence of white noise ($D>0$), from the Langevin equation~{{(\ref{replace})}} we obtain the Fokker-Plank equation:
\begin{equation}
\dot{\rho}{\left ( {\phi,t;\omega,{\overline{h}}} \right )}=-\frac{\partial}{\partial \phi}{\left\{ {{\left ( {\omega-{\mbox{Re}}{\left ( {{\overline{h}}} \right )}\sin\phi+{\mbox{Im}}{\left ( {{\overline{h}}} \right )}\cos\phi} \right )}\rho} \right \}} +D\frac{\partial^2\rho}{\partial \phi^2},{\label{fokkerplank}}
\end{equation}
where $\rho{\left ( {\phi,t;\omega,{\overline{h}}} \right )}$ is the probability distribution of phase $\phi\in [0,2\pi)$ at time $t$, and periodic boundary conditions $\rho{\left ( {0,t;\omega,{\overline{h}}} \right )}=\rho{\left ( {2\pi,t;\omega,{\overline{h}}} \right )},d\rho{\left ( {0,t;\omega,{\overline{h}}} \right )}/d\phi=d
\rho{\left ( {2\pi,t;\omega,{\overline{h}}} \right )}/d\phi$ are imposed.
Since we are concerned with the probability distribution $\rho_{eq}{\left ( {\phi;\omega,r} \right )}$ attained after a long time, we set $\dot{\rho}{\left ( {\phi,t;\omega,r} \right )}=0$ to obtain\cite{sakaguchi,arenas}
\begin{equation}
\rho_{eq}{\left ( {\phi;\omega,r} \right )}=\frac{I(\phi)}{\int_0^{2\pi}I(\phi)d\phi}{\label{stationary}}
\end{equation}
with
\begin{equation}
I(\phi)=\exp{\left ( {\tilde{r}\cos\phi} \right )}\int_0^{2\pi}\exp{\left\{ {-\tilde{\omega}\varphi-\tilde{r}\cos{\left ( {\varphi+\phi} \right )}} \right \}}d\varphi,{\label{density}}
\end{equation}
where $\tilde{\omega}=\omega/D$ and $\tilde{r}=r/D$.
Noting the Ergodic property on the Fokker Plank {{{Eq.}~{(\ref{fokkerplank})}}}, which holds when ${\overline{h}}$ is viewed as a given parameter, we obtain the time average of $z$ by computing the average over the equilibrium distribution $\rho_{eq}{\left ( {\phi;\omega,r} \right )}$:
\begin{equation}
f(\omega,r,r)=\frac{\int_0^{2\pi}e^{i\phi}I(\phi)d\phi}{\int_0^{2\pi}I(\phi)d\phi}.{\label{withd}}
\end{equation}

The transfer function we have obtained here can be considered to be an analogue of the so called TAP equation by the naive mean field model, because the time average of the output $\overline{z}$ is represented as a function of time-averaged local fields $\overline{h}$.
We note, however, that the genuine TAP equation, which is defined for systems with an energy function, should describe the functional relation between the time averaged output and the effective local field that differs in general from the time-averaged one.
The difference between the two types of local fields is known to be the Onsager reaction term in the theory of random spin systems.
We assume that the TAP-like equation may hold even for the present system without an energy function, and suppose it to be given by the following equations:
\begin{eqnarray}
\overline{z_i}&=&f{\left ( {\omega_i,h_i^{{\mbox{\scriptsize TAP}}},h_i^{{\mbox{\scriptsize TAP}}\ast}} \right )},{\label{tapfixedpoint}}{\label{TAPFIXEDPOINT}}\\
h_i^{{\mbox{\scriptsize TAP}}}&=&\overline{h_i}+\gamma^{{\mbox{\scriptsize TAP}}} \overline{z_i}\nonumber\\
&=&\sum_{j\ne i} c_{ij}\overline{z_j}+{\gamma^{{\mbox{\scriptsize TAP}}}}\overline{z_i}\nonumber\\
&=&\sum_\mu\xi_i^\mu \overline{m^\mu}-\alpha \overline{z_i}+{\gamma^{{\mbox{\scriptsize TAP}}}}\overline{z_i}.{\label{taplocalfield}}{\label{TAPLOCALFIELD}}
\end{eqnarray}
In the case with {$\omega_i=0\ (i=1,\ldots,N)$} , by evaluating the free energy of the system,  we can derive an explicit expression of the coefficient ${\gamma^{{\mbox{\scriptsize TAP}}}}$ taking the form (see appendix~\ref{tap}):
\begin{equation}
{\gamma^{{\mbox{\scriptsize TAP}}}}=-\alpha\frac{{\left ( {1-q} \right )}/2D}{1-{\left ( {1-q} \right )}/2D},{\label{gtap}}
\end{equation}
where
\begin{equation}
q=\frac{1}{N}\sum_i |\overline{z_i}|^2.{\label{q}}
\end{equation}
It will, however, be difficult to rigorously derive an expression of ${\gamma^{{\mbox{\scriptsize TAP}}}}$ for the general case with a natural frequency distribution.
Thus we are led to make an assumption that the legitimate expression~{{(\ref{gtap})}} for the case with {$\omega_i=0\ (i=1,\ldots,N)$}\ can naturally be extended to  the general case.
Describing the desired ${\gamma^{{\mbox{\scriptsize TAP}}}}$ requires the introduction of the order parameter $u$ that appears in the SCSNA.

\subsection{Self-Consistent Signal-to-Noise Analysis}

We consider the case with $m^1={\cal O}(1)$ and $m^\mu={\cal O}{\left ( {1/\sqrt{N}} \right )}\ (\mu=2,\ldots,P)$, where we choose pattern~1 as the target.
Assuming, without loss of generality, that $\xi_i^1=1$ for all $i$ and $m^1$ is real owing to the rotational symmetry, the local field {{{Eq.}~{(\ref{taplocalfield})}}} is rewritten in the form:
\begin{equation}
h_i=m^1+\frac{1}{N}{\sum_{\mu> 1}\sum_{j\ne i}\xi_{i}^\mu{{{\xi_{j}^\mu}}^\ast}} s_j+{\gamma^{{\mbox{\scriptsize TAP}}}} s_i,{\label{plainlocalfield}}
\end{equation}
where we have used $s_i,h_i,$ and $m^\mu$ to represent respectively $\overline{z_i},\overline{h_i},$ and $\overline{m^\mu}$ for brevity.

When we consider the case with a finite number of stored patterns, the analysis is straightforward since we already know the form of the transfer functions {{(\ref{nod})}} ($D=0$) and {{(\ref{withd})}} ($D\neq 0$).
Since ${\gamma^{{\mbox{\scriptsize TAP}}}}=0$ and $h_i=m^1$ in this case, we have in the limit $N\rightarrow\infty$ 
\begin{equation}
m^1=\frac{1}{N}\sum_i\xi_i^{1\ast} s_i\rightarrow{{\Bigg \langle}{{f{\left ( {{\omega,m^1,m^1}} \right )}}}{\Bigg \rangle}_\omega},{\label{finite}}
\end{equation}
where $\displaystyle{{\Bigg \langle}{\ldots}{\Bigg \rangle}_\omega}=\int g(\omega)\ldots d\omega$.
Solving this equation numerically, the size of the overlap is evaluated as a function of various parameters including $D$.

In the case of an extensive number of stored patterns ($\alpha>0$), however, cross-talk noise (the second term of {{{Eq.}~{(\ref{plainlocalfield})}}}) in the local fields becomes to an appreciable extent, then one has to employ the method of SCSNA.
The crux of the SCSNA is the evaluation of the variance of cross-talk noise that interferes occurrence of retrieval state.

According to the prescription of the SCSNA\cite{yoshioka2}, the local fields {{(\ref{plainlocalfield})}} is assumed to be in the form:
\begin{equation}
h_i=m^1+\frac{\lambda}{N}{\sum_{\mu> 1}\sum_{j\ne i}\xi_{i}^\mu{{{\xi_{j}^\mu}}^\ast}} s_j^\mu+{\gamma^{\mbox{\scriptsize SCSNA}}} s_i+{\gamma^{{\mbox{\scriptsize TAP}}}} s_i.{\label{assumption}}
\end{equation}
where $s_j^\mu$, whose explicit expression is given later, is a quantity that is very close to $s_i^\mu$ and has negligible correlation in the limit $N\rightarrow\infty$, and $\lambda$ and ${\gamma^{\mbox{\scriptsize SCSNA}}}$ are to be self-consistently determined in the course of analysis.

Defining $\tilde{h_i}=m^1+\frac{\lambda}{N}{\sum_{\mu> 1}\sum_{j\ne i}\xi_{i}^\mu{{{\xi_{j}^\mu}}^\ast}} s_j^\mu$, {{{Eq.}~{(\ref{tapfixedpoint})}}} reads   
\begin{equation}
s_i=f{\left ( {\omega_i,\tilde{h_i}+{\gamma^{{\mbox{\scriptsize TOTAL}}}} s_i,{\left ( {\tilde{h_i}+{\gamma^{{\mbox{\scriptsize TOTAL}}}} s_i} \right )}^\ast} \right )},{\label{fixedpoint}}
\end{equation}
where ${\gamma^{{\mbox{\scriptsize TOTAL}}}}={\gamma^{\mbox{\scriptsize SCSNA}}}+{\gamma^{{\mbox{\scriptsize TAP}}}}$.

Considering a general case with ${\gamma^{{\mbox{\scriptsize TOTAL}}}}\neq 0$, we solve {{{Eq.}~{(\ref{fixedpoint})}}} for $s_i$ to obtain the renormalized output $s_i=\tilde{f}{\left ( {\omega_i,\tilde{h_i},\tilde{h_i}^\ast} \right )}$.
Performing a Taylor expansion of $\tilde{f}{\left ( {\omega_i,\tilde{h_i},\tilde{h_i}^\ast} \right )}$ about ${\left ( {\tilde{h_i^\mu},\tilde{h_i^\mu}^\ast} \right )}$, we have
\begin{equation}
s_i=s_i^\mu+{\left.\frac{\partial \tilde{f}}{\partial \tilde{h}}\right |_{{{\left ( {{\tilde{h_{{{i}}}^{{{\mu}}}},\tilde{h_{{{i}}}^{{{\mu}}}}^\ast}} \right )}}}}\frac{\lambda}{N}\sum_{j\ne i}\xi_i^\mu{{{\xi_j^\mu}}^\ast}s_j^\mu+{\left.\frac{\partial \tilde{f}}{\partial \tilde{{{{h}}^\ast}}}\right |_{{{\left ( {{\tilde{h_{{{i}}}^{{{\mu}}}},\tilde{h_{{{i}}}^{{{\mu}}}}^\ast}} \right )}}}}{{{{\left ( {\frac{\lambda}{N}\sum_{j\ne i}\xi_i^\mu{{{\xi_j^\mu}}^\ast}s_j^\mu} \right )}}}^\ast}{\label{taylor}}
\end{equation}
with
\begin{eqnarray}
\tilde{h_i^\mu}&=&m^1+\frac{\lambda}{N}\sum_{\nu \ne 1,\mu}\sum_{j\ne i}\xi_i^\nu{{{\xi_j^\nu}}^\ast}s_j^\nu,\\
s_i^\mu&=&\tilde{f}{\left ( {\omega_i,\tilde{h_i^\mu},\tilde{h_i^\mu}^\ast} \right )}.
\end{eqnarray}
Substituting {{{Eq.}~{(\ref{taylor})}}} into {{{Eq.}~{(\ref{plainlocalfield})}}} and comparing the result with {{{Eq.}~{(\ref{assumption})}}} (see appendix~\ref{evaluation} for details) we obtain
\begin{eqnarray}
&&m^1+\frac{\lambda}{N}{\sum_{\mu> 1}\sum_{j\ne i}\xi_{i}^\mu{{{\xi_{j}^\mu}}^\ast}} s_j^\mu+{\gamma^{\mbox{\scriptsize SCSNA}}} s_i+{\gamma^{{\mbox{\scriptsize TAP}}}} s_i\nonumber\\
&=&m^1+\frac{1+u\lambda}{N}\sum_{\mu>1}\sum_{j\ne i}\xi_i^\mu{{{\xi_j^\mu}}^\ast}s_j^\mu+\alpha u \lambda s_i+{\gamma^{{\mbox{\scriptsize TAP}}}} s_i {\label{compare}}{\label{COMPARE}}
\end{eqnarray}
where
\begin{equation}
u=\frac{1}{N}\sum_i{\left.\frac{\partial \tilde{f}}{\partial \tilde{h}}\right |_{{\left ( {\tilde{h_i},\tilde{h_i}^\ast} \right )}}}.{\label{u}}
\end{equation}
Since this equation holds for every site $i$, $\lambda$ and ${\gamma^{\mbox{\scriptsize SCSNA}}}$ are self-consistently determined as
\begin{eqnarray}
\lambda&=&\frac{1}{1-u}, {\label{lambda}}\\
{\gamma^{\mbox{\scriptsize SCSNA}}}&=&\alpha\frac{u}{1-u} {\label{gamma}}.
\end{eqnarray}

The variance of renormalized noise $\tilde{N_i}=\frac{\lambda}{N}{\sum_{\mu> 1}\sum_{j\ne i}\xi_{i}^\mu{{{\xi_{j}^\mu}}^\ast}} s_j^\mu$ in {the right hand side}\ of {{{Eq.}~{(\ref{assumption})}}} can be evaluated by noting that ${\mbox{Re}}{\left ( {\tilde{N_i}} \right )}$ and ${\mbox{Im}}{\left ( {\tilde{N_i}} \right )}$ distribute over sites obeying an identical  Gaussian distribution independently and the site average of $|\tilde{N_i}|^2$ can be replaced by the pattern average: one has
\begin{equation}
\frac{1}{N}\sum_i|\tilde{N_i}|^2={{\Bigg \langle}{|\tilde{N_i}|^2}{\Bigg \rangle}_\xi}=\alpha |\lambda|^2q.{\label{variance}}
\end{equation}
To represent the noise as $\tilde{N_i}=\sqrt{\alpha r}{\left ( {x_i+i y_i} \right )}/2$, where $x_i$ and $y_i$ are real and obey a normal Gaussian, we define
\begin{equation}
r=2|\lambda|^2q=\frac{2q}{\left|1-u\right|^2}.{\label{r}}
\end{equation}

Summarizing {{{Eqs.}~{(\ref{overlap})}}},{{(\ref{q})}},{{(\ref{assumption})}},{{(\ref{fixedpoint})}},{{(\ref{u})}},{{(\ref{lambda})}},{{(\ref{gamma})}}, and {{(\ref{r})}}, we have a set of macroscopic order parameter equations 
\begin{eqnarray}
s&=&f{\left ( {\omega,\tilde{h}+{\gamma^{{\mbox{\scriptsize TOTAL}}}} s,{\left ( {\tilde{h}+{\gamma^{{\mbox{\scriptsize TOTAL}}}} s} \right )}^\ast} \right )}, {\label{sctrf}}\\
\tilde{h}&=&m+\frac{\sqrt{\alpha r}}{2}{\left ( {x+i y} \right )},{\label{sch}} \\
{\gamma^{{\mbox{\scriptsize TOTAL}}}}&=&{\gamma^{\mbox{\scriptsize SCSNA}}}+{\gamma^{{\mbox{\scriptsize TAP}}}},{\label{scsum}}\\
m&=&{{\Bigg \langle}{\Bigg \langle}{{\tilde{f}{\left ( {{\omega,\tilde{h},\tilde{h}^\ast}} \right )}}}{\Bigg \rangle}{\Bigg \rangle}}, {\label{scm}}\\
q&=&{{\Bigg \langle}{\Bigg \langle}{\left|{\tilde{f}{\left ( {{\omega,\tilde{h},\tilde{h}^\ast}} \right )}}\right|^2}{\Bigg \rangle}{\Bigg \rangle}},{\label{scq}} \\
\sqrt{\alpha r}u&=&{{\Bigg \langle}{\Bigg \langle}{{\left ( {x-i y} \right )}{\tilde{f}{\left ( {{\omega,\tilde{h},\tilde{h}^\ast}} \right )}}}{\Bigg \rangle}{\Bigg \rangle}}, {\label{scu}}{\label{SCU}}\\
r&=&\frac{2q}{{\left ( {1-u} \right )}^2},{\label{scr}}{\label{SCR}} \\
{\gamma^{\mbox{\scriptsize SCSNA}}}&=&\alpha\frac{u}{1-u},{\label{scgamma}}
\end{eqnarray}
where ${{\Bigg \langle}{\Bigg \langle}{\ldots}{\Bigg \rangle}{\Bigg \rangle}}$ represents $\displaystyle{\Bigg \langle}{{\Bigg \langle}{\ldots}{\Bigg \rangle}_\omega}{\Bigg \rangle}_{\tilde{h}}=\int \frac{g{\left ( {\omega} \right )}}{2\pi}\exp{\left ( {-\frac{x^2+y^2}{2}} \right )}\ldots d\omega dx dy$.
Detailed derivation of {{{Eqs.}~{(\ref{scu})}}} and {{(\ref{scr})}} is given in appendix~\ref{miscellaneous}.

To discuss the generalized expression for ${\gamma^{{\mbox{\scriptsize TAP}}}}$ for the case with a distribution of natural frequencies, we consider for the moment the case with {$\omega_i=0\ (i=1,\ldots,N)$} , where {{{Eq.}~{(\ref{gtap})}}} exactly holds.
In this case it turns out that ${\gamma^{{\mbox{\scriptsize TOTAL}}}}=0$ by a rough argument given below.
Note that ${\gamma^{{\mbox{\scriptsize TOTAL}}}}=0$ implies ${\tilde{f}{\left ( {{\omega,\tilde{h},{{{\tilde{h}}}^\ast}}} \right )}}={f{\left ( {{\omega,h,{{{h}}^\ast}}} \right )}}$, and that the effective transfer function~{{(\ref{withd})}} becomes
\begin{equation}
{f{\left ( {{0,r,r}} \right )}}=\frac{\int_0^{2\pi}\cos\phi\exp{\left ( {\tilde{r}\cos\phi} \right )} d\phi}{\int_0^{2\pi}\exp{\left ( {\tilde{r}\cos\phi} \right )} d\phi}.{\label{bessel}}
\end{equation}
Using {{{Eqs.}~{(\ref{circulate})}}} and {{(\ref{bessel})}} and performing the average over the Gaussian distribution with unit variance for {{{Eq.}~{(\ref{scu})}}}, we obtain
\begin{equation}
u={\left ( {1-q} \right )}/2D.{\label{special}}
\end{equation}
Then, from {{{Eq.}~{(\ref{gtap})}}}, we have
\begin{equation}
{\gamma^{{\mbox{\scriptsize TAP}}}}=-\alpha\frac{u}{1-u},{\label{generalgtap}}
\end{equation}
which immediately reconfirms
\begin{equation}
{\gamma^{{\mbox{\scriptsize TOTAL}}}}={\gamma^{\mbox{\scriptsize SCSNA}}}+{\gamma^{{\mbox{\scriptsize TAP}}}}=0.
\end{equation}

Now we observe that {\it in the case with {$\omega_i=0\ (i=1,\ldots,N)$}\ the Onsager reaction term ${\gamma^{{\mbox{\scriptsize TAP}}}} s_i$ cancels out with the term ${\gamma^{\mbox{\scriptsize SCSNA}}} s_i$ that emerges as a result of the evaluation of the correlation between the state of oscillators and the stored patterns.}
Then {\it we assume {{{Eq.}~{(\ref{generalgtap})}}} to hold generally so that ${\gamma^{{\mbox{\scriptsize TOTAL}}}}=0$.}
As will be shown later the results obtained based on this assumption show good agreement with the results of numerical simulations.

\section{Behaviors of the network with a discrete natural frequency distribution}{\label{application}}

For the sake of simplicity we focus on the behavior of the oscillator network with a discrete natural frequency distribution $g{\left ( {\omega} \right )}$:
\begin{equation}
g(\omega)=\frac{1-a}{2}\delta{\left ( {\omega+\omega_1} \right )}+a\delta{\left ( {\omega} \right )}+\frac{1-a}{2}\delta{\left ( {\omega-\omega_1} \right )},{\label{peaks}}
\end{equation}
where $a$ represents the ratio of the number of oscillators with $\omega_i=0$ to the total number of oscillators $N$.

\subsection{Appearance of a window for break down of the retrieval states}{\label{window}}

\begin{figure}
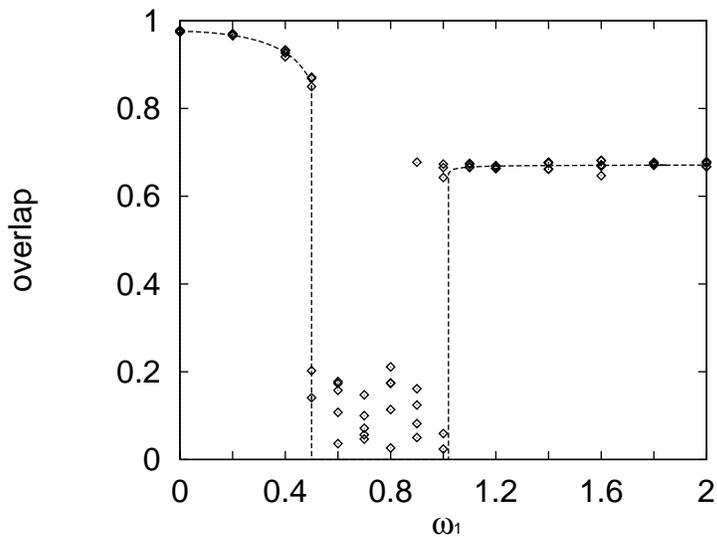

{}{}
\caption{$\omega_1$-dependence of the overlap obtained from the present analysis is plotted together with the results of numerical simulations with $N=4000$ for the case with $\alpha=0.02,a=0.7,D=0$.
Since theoretical analysis is based on taking a time average of physical quantites of interest, the results of simulation are displayed in terms of time-averaged quantities $\overline{\left|m^\mu\right|}=\overline{\left |\frac{1}{N}\sum_i\xi_i^\mu z_i\right |}$.}
{\label{break}}
\end{figure}

To roughly sketch the effects of the natural frequency distribution with three frequency components, we investigate the behavior of the overlap with change of $\omega_1$ in the case of $D=0$.
In {{{Fig.}~\ref{break}}}, we give $\omega_1$-dependence of the overlap calculated from {{{Eqs.}~{(\ref{sctrf})}}}-{{(\ref{scgamma})}}  and the result of numerical simulations with $N=4000$ for the case with  $\alpha=0.02,a=0.7,$ and $D=0$.
Good agreement between the theory and numerical simulations implies the validity of the present analysis.

As is expected, an entrainment indeed occurs in the case of small $\omega_1$ ($\omega_1{\lesssim}0.5$), where one has successful retrieval accompanying a large size of overlap.
Even in the case of large $\omega_1$ ($1.0{\lesssim}\omega_1$) successful retrieval can be realized with small size of overlap, since natural frequency of $aN$ oscillators remains 0.

To give an qualitative explanation for the occurrence of a window for break down of the retrieval states, we consider the case with $\alpha=0$ for the moment.
We can easily obtain the values of order parameters $m,q,$ and $u$ as a function of $\omega_1$ (see appendix~\ref{a0}) as shown in {{{Fig.}~\ref{reason}}}{{\bf (a)}} .
We see a phase transition to occur at $\omega_1\approx0.7$, and $u$ is seen to increase as $\omega_1$ approaches $\omega_1^c$: $u\rightarrow 1$ as $\omega_1\rightarrow\omega_1^c$, while $q=1$ for $\omega_1<\omega_1^c$ owing to the entrainment.

Even when $\alpha\neq0$, such an enhancement of $u$ around $\omega_1^c$ remains unchanged as can be seen in {{{Fig.}~\ref{reason}}}{{\bf (b)}}.
Noting {{{Eq.}~{(\ref{scr})}}} we can easily understand the noise variance $\alpha r/2=\alpha q/{\left ( {1-u} \right )}^2$ may be enhanced accordingly in the interval $0.5{\lesssim}\omega_1{\lesssim} 1.0$, where retrieval states disappears.

\begin{figure}
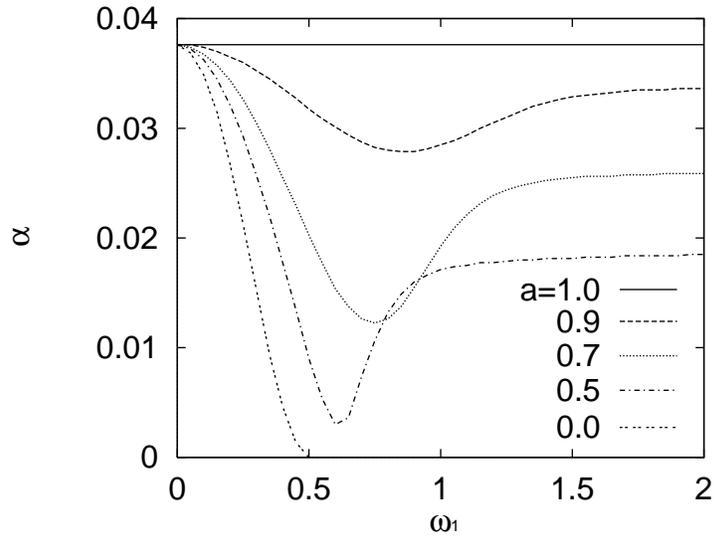

{}{}
\caption{$\omega_1-\alpha$ phase diagram representing the behavior of strage capacities for the various values of $a$ in the case of $D=0$.}
{\label{breakpd}}
\end{figure}

In {{{Fig.}~\ref{breakpd}}} we draw the $\omega-\alpha$ phase diagram to show the behavior of the storage capacity as a function of $\omega_1$.
The window observed in {{{Fig.}~\ref{break}}} turns to arise from the valley of $\alpha_c{\left ( {\omega_1} \right )}$ curves.

\subsection{Effect of white noise}

Behaviors of synchronization in the networks of coupled oscillators with white noise differ from those of the deterministic networks with $D=0$.
In the absence of white noise ($D=0$), one can in general divide the oscillators with $\omega_i\neq 0$ into two groups of synchronized and desynchronized oscillators accordingly to the criterion of whether the phase velocity of an oscillation vanishes or not.
Noting {{{Eq.}~{(\ref{scm})}}} and the form of effective transfer function with $D=0$ ({{{Eq.}~{(\ref{nod})}}}) illustrated in {{{Fig.}~\ref{field}}}, we  find that desynchronized oscillators do not contribute to the value of overlap $m$, though they contribute to the value of other order parameters such as $q$ and $u$.
This is because we are concerned with the time averaged behavior of the local fields, where the time-averaged phase difference between the desynchronized oscillators with natural frequencies $\omega$ and $-\omega$ should be $\pi$ in the absence of white noise ({i e. }\ ${f{\left ( {{-\omega,h,h^\ast}} \right )}}=-{f{\left ( {{\omega,h,h^\ast}} \right )}}$ for $|h|<\omega$).
In the case with white noise ($D\neq 0$), however, the phase of each oscillator with $\omega_i\neq 0$ evolves with a certain non-zero time-averaged phase velocity, since the action of white noise prevents any oscillators from settling into fixed points.
Hence it becomes impossible to distinguish between the synchronized and desynchronized oscillators any more.
Nevertheless a finite value of the overlap is realized because of the existence of the equilibrium probability distribution on $[0,2\pi)$ that is achieved after a long time with $\overline{z}\neq 0$.

\begin{figure}
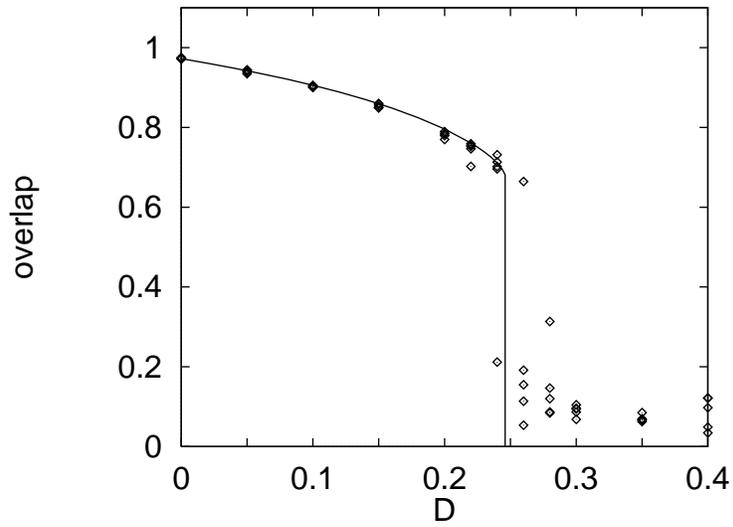

{}{}
\caption{$D$-dependence of the overlap $m$ obtained from the present analysis is plotted together with the results of numerical simulations with $N=4000$ in the case with $a=0.7,\alpha=0.01,\omega_1=0.3$.}
{\label{intensity}}
\end{figure}

In {{{Fig.}~\ref{intensity}}} we display the behavior of the overlap as a function of the noise intensity $D$ obtained by the theoretical analysis and numerical simulations.
As expected, the size of the overlap decreases as the noise intensity increases until the system undergoes a discontinuous transition at a critical noise intensity $D^c$, above which a disordered state with $m=0$ is realized.
Good agreement between the two results implies the validity of our  treat ment based on the time averaged local fields together with the assumption that the TAP-like equation also holds in the case with a distribution of natural frequencies.

\begin{figure}
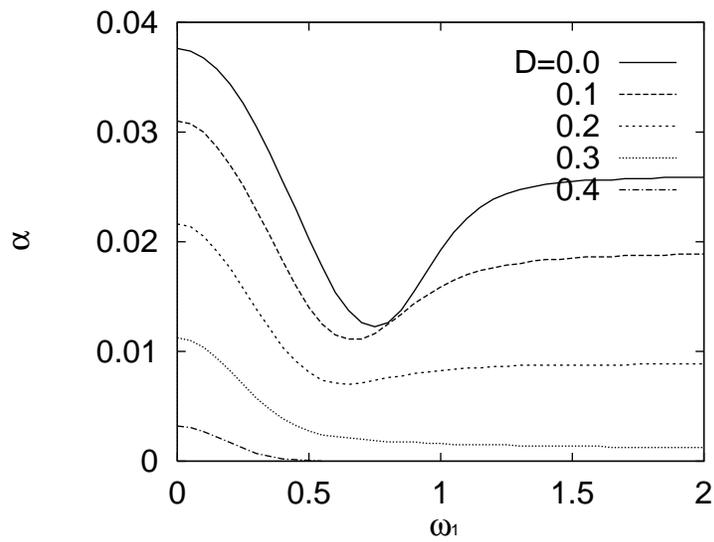

{}{}
\caption{Same as in {{{Fig.}~\ref{breakpd}}} for various values of $D$ in the case of $a=0.7$.}
{\label{dpd}}
\end{figure}

In {{{Fig.}~\ref{dpd}}}, we give $\omega_1-\alpha$ phase-diagram that represents the storage capacity plotted as a function of $\omega_1$ for various values of $D$.
In most of the region of $\omega_1$, the storage capacity $\alpha_c$ decreases as the noise intensity increases.
We see that for $D$ smaller than a certain critical $D_0$, $\alpha_c{\left ( {\omega_1} \right )}$ exhibits non-monotonic behavior with change in $\omega_1$, while for $D>D_0$ $\alpha_c{\left ( {\omega_1} \right )}$ is monotonically decreasing with $\omega_1$.
The occurrence of a window for the break down of the retrieval states with fixed $\alpha$ for $D<D_0$ turns out to be attributed to the appearance of valley caused by the non-monotonic behavior of the $\alpha_c{\left ( {\omega_1} \right )}$ curve as in the case of $D=0$ ({{{Fig.}~\ref{break}}}). 

\section{Summary and discussions}{\label{discussion}}

We have investigated properties of an associative memory model of oscillator neural networks based on simple phase oscillators, where the influence of white noise together with a natural frequency distribution is considered in the case of an extensive number of stored patterns.
In the presence of white noise every oscillator as well as its local field undergoes fluctuating motions even in the stationary state after a long time.
To deal with such a situation we have taken an approach based on the concept of the TAP-like equation.
To approximately derive the TAP-like equation for the system without an energy function we have taken the time average for the fluctuating local field of each oscillator neuron to make it constant in time.
On the basis of the time-averaged local field we have dealt with the single-body Fokker-Plank equation to obtain the time averaged outputs of the oscillators in the stationary state, from which we have evaluated the effective transfer function.
The relation between the time-averaged output and the local fields  involving such a transfer function can be viewed as an analogue of the naive TAP-like equation without considering the so called Onsager reaction term.
We have supposed the proper form of TAP-like equation to be given by appropriately adjusting the Onsager reaction term such that setting {$\omega_i=0\ (i=1,\ldots,N)$}\ naturally leads to the legitimate TAP equation, which we have obtained from the evaluation of the Gibbs free energy.
Applying the SCSNA to this TAP-like equation, we have obtained the macroscopic order parameter equations, based on which the properties of associative memory of the network has been studied.

Assuming a discrete symmetric natural frequency distribution with three frequency components for the sake of simplicity, we have presented the phase diagram showing the behavior of storage capacity as a function of the parameter $\omega_1$ representing the width of the natural frequency distribution.
In the case of $D=0$ the storage capacity $\alpha_c$ has been found to exhibit non-monotonic behavior as $\omega_1$ is varied and to attain a  minimum at a certain $\omega_1$.
As a result of the occurrence of the valley in the $\omega_1-\alpha_c$ curve the break down of the retrieval state with fixed $\alpha$ occurs for intermediate values of $\omega_1$.
When noise is present such a behavior has been found to be somewhat relaxed, and only for small values of the noise intensity $D$ the phenomenon of the break down of the retrieval state can be observed.
Our analytical result has shown excellent agreement with the results of numerical simulations.

Our results show that associative memory based on temporal coding can be realized in the network of simple phase oscillators even in the presence of not only a distribution of natural frequencies but also external white noise.
The result that temporal coding is robust against the existence of environmental noise is remarkable.
Memory retrieval occurs in such a way that the oscillators undergo synchronized motions with the phase difference $\phi_i-\phi_j$ between any two of the oscillators $i$ and $j$ settling into, for long times, the difference $\theta_i^\mu-\theta_j^\mu$ of the memory pattern~$\mu$.
In our model $\theta_i^\mu$ is chosen from uniformly distributed random numbers on $[0,2\pi)$.
The resultant behavior, however, is qualitatively the same as that for our previous work ($D=0$)\cite{yamana} on the special case where $\theta_i^\mu=0$ or $\pi$ and hence $J_{ij}$ is given by the well-known form $J_{ij}=\frac{1}{N}\sum_\mu\xi_i^\mu\xi_j^\mu$ with $\xi_i^\mu=\pm 1$.
A characteristic feature of memory retrieval accompanying synchronization is that, in contrast to networks with fixed point type attractors, each neuron exhibit oscillations in the local field or the membrane potential that are easily detected by other neurons in a certain network to determine whether memory retrieval is successful or not.
Also worth noting is the appearance of two types of retrieval states with respect to the degree of synchronization: a high degree of synchronization that occurs for small $\omega_1$ with overlap $m$ large and a low degree of synchronization that occurs for large $\omega_1$ with $m$ small.

The fundamental assumption we have used in the present study is the existence of the TAP-like equation {{(\ref{tapfixedpoint})}} and {{(\ref{taplocalfield})}} for our system together with the expression of ${\gamma^{{\mbox{\scriptsize TAP}}}}$({{{Eq.}~{(\ref{generalgtap})}}}).
In the case of {$\omega_i=0\ (i=1,\ldots,N)$}\ there occurs no problem because the genuine TAP equation exists as has been shown.
In this case we have found that ${\gamma^{{\mbox{\scriptsize TAP}}}}$ ({{{Eq.}~{(\ref{gtap})}}}) is exactly canceled out by ${\gamma^{\mbox{\scriptsize SCSNA}}}$, as in the case of the network of AGS, to yield ${\gamma^{{\mbox{\scriptsize TOTAL}}}}=0$ together with the order parameter equations, that recover the ones by Cook\cite{cook}, who analyzed $Q$-state spin model including the case with $Q\rightarrow\infty$ for arbitrary temperatures by means of the replica symmetric approximation.

In the case with distributed natural frequencies, to obtain the form of the effective transfer function of the TAP-like equation, we  replace time-dependent local fields $h_i$ in {{{Eq.}~{(\ref{nativedynamics})}}} by their time-averaged ones, and assume that the Onsager reaction term of the form: ${\gamma^{{\mbox{\scriptsize TAP}}}} s_i=-\alpha u/{\left ( {1-u} \right )} s_i$ appears in effective local field as a result of fluctuation of local fields.
The assumption of this form of the Onsager reaction term  yields ${\gamma^{{\mbox{\scriptsize TOTAL}}}}=0$, which leads to ${\tilde{f}{\left ( {{\omega,h,h^\ast}} \right )}}={f{\left ( {{\omega,h,h^\ast}} \right )}}$ in {{{Eqs.}~{(\ref{scm})}}}-{{(\ref{scu})}}.

Meanwhile the theoretical result we have obtained here is recovered by the TAP-like equation evaluated by simply replacing the local fields $h_i$ in {{{Eq.}~{(\ref{nativedynamics})}}} by the time-dependent local field of the form:
\begin{equation}
 h_i=\tilde{h_i}+{\gamma^{\mbox{\scriptsize SCSNA}}} z_i, \label{newassumption}
\end{equation}
where the renormalized local field: $\tilde{h_i}=\xi_i^1 m+{\left ( {1/N} \right )}\sum_{\mu>1}\sum_j \xi_i^\mu\xi_j^{\mu\ast} s_j^\mu$ is constant in time.
Note that, by use of this assumption, we can derive the order parameter equations without knowing the form of ${\gamma^{{\mbox{\scriptsize TAP}}}}$.
In the present case both procedures give the same result, however, in some cases it seems that the later procedure gives the more accurate form of the transfer function.
We will discuss details on this point somewhere else.

Some special cases of the present model have also been investigated by several authors other than Cook.
Arenas {{ et al.}}\ \cite{arenas} have investigated the case with $\alpha=0$, where the natural frequency distribution is assumed to obey a Gaussian distribution.
The result of this case can also be recovered by the present analysis.

To our knowledge the case with distributed natural frequencies and $\alpha>0$ was first studied by Park {{ et al.}}\ \cite{park} for different synaptic couplings by means of replica calculations based on the  energy that is defined so as to satisfy $\dot{\phi_i}=-d{H_{{\mbox{\scriptsize Park}}}}/d\phi_i+\eta_i$.
In our case the $H_{{\mbox{\scriptsize Park}}}$ takes the form $H_{{\mbox{\scriptsize Park}}}=-\frac{1}{2}\sum_{i\neq j}J_{ij}\cos{\left ( {\phi_i-\phi_j-\beta_{ij}} \right )}-\sum_i\omega_i\phi_i$.
However this energy does not make any sense because the equilibrium distribution $\exp{\left ( {-H_{{\mbox{\scriptsize Park}}}/D} \right )}$ does not satisfy the periodical boundary condition $P{\left ( {{\left\{ {\phi_i} \right \}}} \right )}=P{\left ( {{\left\{ {\phi_i+2\pi} \right \}}} \right )}$.
		
Aonishi {{ et al.}}\ \cite{aonishi2} studied the case with $D=0$ and a Gaussian distribution for natural frequencies by considering that $q=1$ holds in the set of SCSNA eqs. based on a different scheme from ours even in the presence of the group of the desynchronized oscillators.

Yamana {{ et al.}}\ also studied the deterministic oscillator network ($D=0$) with a  discrete distribution of natural frequencies that stores binary patterns by making an approximation that the motions of the group of desynchronized oscillators do not exert an influence on the behavior of the synchronized oscillators.
Discarding the effect of desynchronized oscillators corresponds to considering the transfer function that takes the value zero inside the circle with radius $\omega$ (see {{{Fig.}~\ref{field}}}).
For a wide class of natural frequency distributions this scheme seems to work to a good approximation in the case with $D=0$, because the contribution of the desynchronized oscillators to such order parameters as $m,q,$ and $u$ is small.
It is noted, however, that, in the case of $D\neq 0$, the phase of every oscillator with $\omega_i\neq 0$ evolves with a certain non-zero time-averaged velocity and hence one cannot distinguish between synchronized and desynchronized oscillators.
Accordingly for stochastic networks with $D\neq 0$ methods based on neglecting the effect of the desynchronized oscillators will not make sense and one has to deal with all of the oscillators equally as in the present analysis.

Finally, we briefly discuss the relevance of our results to biologically related models of associative memory.
Biologically relevant models\cite{bressloff1,bressloff2,xiao,terman,gabbay,ernst} should be based on  such  spiking neurons  as the Hodgkin-Huxley type\cite{xiao} and integrate-and-fire type neurons\cite{bressloff1,bressloff2,gabbay,ernst}.
A simple  integrate-and-fire neuron that is defined by 1-dimensional linear equation except for firing event can be described in terms of phase that is obtained by properly scaling the 1-dimensional output variable.
Synaptic couplings implimented into spiking neural networks are often assumed to incorporate the so called alpha function\cite{gabbay} or its variant represented by the dynamics of a certain gating variable\cite{xiao,terman}.
So, major differences between the simple phase oscillator model we have dealt with  on the basis of the diffusive couplings among the oscillators and the spiking model will be the form of the synaptic couplings together with symmetry of an individual oscillator with respect to rotation of the phase variable.
While the present model is assumed to take a sinusoidal phase interaction for simplicity, a spiking model with a synaptic interaction based on the alpha function takes the form of pulse like couplings\cite{bressloff1,bressloff2,gabbay,ernst}, which  will lead to considering higher harmonics in the phase interaction.

A spiking neural network model of  associative memory we previously studied using FitzHugh-Nagumo neurons exhibits a nearly comparable size of the storage capacity to that of the standard analog network with the transfer function $F(h)={\left ( {\mbox{sgn}{\left ( {h} \right )}+1} \right )}/2$ that is larger than the storage capacity of the present model\cite{yoshioka}.
It will then be of interest  to observe the outcome of introducing higher harmonics in the phase interaction of the simple phase oscillator model.
We expect the storage capacity of the phase oscillator network to increase when the higher harmonics is taken into account.
Such analysis is now under way.

The problem of investigating properties of neurons synchronizing the envelope of a burst of spikes is also of interest, but is beyond the scope of the present paper, which aims studying the effects of such heterogeneities as a natural frequency distribution and external noise on the robustness of temporal coding in the oscillator network of associative memory.
We consider that taking not only phase but also amplitude as variables for oscillatory neurons will provide a solvable model suitable for studying the case with such synchronization in networks of bursting neurons, which is also under way.
 
\section*{Acknowledgment}

One of the authors (M. Yoshioka) would like to acknowledge the support Grant-in-Aid for Encouragement of Young Scientists (no. 4415) from the Ministry of Education.

{\appendix}

\section{Derivation of the TAP equation~{{(\ref{tapfixedpoint})}} and {{(\ref{taplocalfield})}} in the case with {$\omega_i=0\ (i=1,\ldots,N)$} }{\label{tap}}

To obtain the TAP equation for the present model with the energy function~{{(\ref{hamiltonian})}} we follow the method of Plefka\cite{plefka} and Nakanishi\cite{nakanishi} used for the SK model and neural networks of Ising spins.

The Hamiltonian~{{(\ref{hamiltonian})}} with a complex-valued external field $R_i+i I_i$ included reads
\begin{eqnarray}
\tilde{H}&=&aH-\sum_i{\left ( {R_i \cos\phi_i+I_i \sin\phi_i} \right )}\nonumber\\
&=&-\frac{a}{2}\sum_{i\neq j}{{{c_{ij}}}^\ast}z_i {{{z_j}}^\ast}-\sum_i{\left ( {R_i \cos\phi_i+I_i \sin\phi_i} \right )},{\label{htilde}}
\end{eqnarray}
where $a$ is introduced for the analysis below.
Applying Legendre transformation to the free energy corresponding to the Hamiltonian $\tilde{H}$, one has
\begin{equation}
{G{\left ( {a,{\left\{ {s_{i}} \right \}}} \right )}}=-\beta^{-1}\log\mbox{Tr}\exp{\left ( {-\beta\tilde{H}} \right )}+\sum_i{\left ( {R_i x_{i}+I_i y_{i}} \right )},{\label{gfunc}}
\end{equation}
where $\beta=1/D$ and $s_{i}=x_{i}+i y_{i}={{\Big \langle}{\cos\phi_i}{\Big \rangle}_a}+i{{\Big \langle}{\sin\phi_i}{\Big \rangle}_a}={{\Big \langle}{z_i}{\Big \rangle}_a}$.
${{\Big \langle}{\ldots}{\Big \rangle}_a}$ denotes expectation with respect to the Hamiltonian $\tilde{H}$.

We perform a Taylor expansion with respect to $a$
\begin{equation}
{G{\left ( {a,{\left\{ {s_{i}} \right \}}} \right )}}=\sum_{n=0}\frac{{G^{({n})}}}{n!} a^n,
\end{equation}
where ${G^{({n})}}=\partial^n G/\partial a^n|_{a=0}$.
Noting $R_i+i I_i=\partial G/\partial x_{i}+i \partial G/\partial y_{i}$, we rewrite {{{Eq.}~{(\ref{gfunc})}}} in the form 
\begin{eqnarray}
&&{G{\left ( {a,{\left\{ {s_{i}} \right \}}} \right )}}\nonumber\\
&=&-\beta^{-1}\log\mbox{Tr}\exp{\left\{ {-a\beta H+\beta\sum_{n=0}\sum_i\frac{a^n}{n!}{\left ( {\frac{\partial {G^{({n})}}}{\partial x_{i}}\cos\phi_i+\frac{\partial {G^{({n})}}}{\partial y_{i}}\sin\phi_i} \right )}} \right \}}\nonumber\\
&&+\sum_{n=0}\sum_i\frac{a^n}{n!}{\left ( {\frac{\partial {G^{({n})}}}{\partial x_{i}}x_{i}+\frac{\partial {G^{({n})}}}{\partial y_{i}}y_{i}} \right )}\nonumber\\
&=&-\beta^{-1}\log Z+\sum_i{\left ( {\frac{\partial {G^{({0})}}}{\partial x_{i}}x_{i}+\frac{\partial {G^{({0})}}}{\partial y_{i}}y_{i}} \right )}\nonumber\\
&&-\beta^{-1}\log\frac{1}{Z}\mbox{Tr}\exp{\left\{ {\beta\sum_i{\left ( {\frac{\partial {G^{({0})}}}{\partial x_{i}}\cos\phi_i+\frac{\partial {G^{({0})}}}{\partial y_{i}}\sin\phi_i} \right )}} \right \}}\nonumber\\
&&\times\exp{\left\{ {-a\beta H+\beta\sum_{n=1}\sum_i\frac{a^n}{n!}{\left ( {\frac{\partial {G^{({n})}}}{\partial x_{i}}\cos\phi_i+\frac{\partial {G^{({n})}}}{\partial y_{i}}\sin\phi_i} \right )}} \right \}}\nonumber\\
&&+\sum_{n=1}\sum_i\frac{a^n}{n!}{\left ( {\frac{\partial {G^{({n})}}}{\partial x_{i}}x_{i}+\frac{\partial {G^{({n})}}}{\partial y_{i}}y_{i}} \right )}\nonumber\\
&=&{G^{({0})}}\nonumber\\
&&-\beta^{-1}\log{{\Big \langle}{\exp{\left\{ {-a\beta H+\beta\sum_{n=1}\sum_i\frac{a^n}{n!}{\left ( {\frac{\partial {G^{({n})}}}{\partial x_{i}}\cos\phi_i+\frac{\partial {G^{({n})}}}{\partial y_{i}}\sin\phi_i} \right )}} \right \}}}{\Big \rangle}_0}\nonumber\\
&&+\sum_{n=1}\sum_i\frac{a^n}{n!}{\left ( {\frac{\partial {G^{({n})}}}{\partial x_{i}}x_{i}+\frac{\partial {G^{({n})}}}{\partial y_{i}}y_{i}} \right )}\nonumber\\
&=&{G^{({0})}}-\beta^{-1}\log{{\Big \langle}{\exp{\left\{ {-a\beta H+\beta\sum_{n=1}\frac{a^n}{n!}A_{n}} \right \}}}{\Big \rangle}_0}{\label{firstrewrite}}
\end{eqnarray}
with
\begin{equation}
A_{n}=\frac{1}{2}\sum_i{\left\{ {{\left ( {\partial {G^{({n})}}/\partial x_{i}+i \partial {G^{({n})}}/\partial y_{i}} \right )}{\left ( {z_i-s_{i}} \right )}^\ast+{\left ( {\partial {G^{({n})}}/\partial x_{i}+i \partial {G^{({n})}}/\partial y_{i}} \right )}^\ast{\left ( {z_i-s_{i}} \right )}} \right \}},
\end{equation}
where $Z=\mbox{Tr}\exp{\left\{ {\beta\sum_i{\left ( {\frac{\partial {G^{({0})}}}{\partial x_{i}}\cos\phi_i+\frac{\partial {G^{({0})}}}{\partial y_{i}}\sin\phi_i} \right )}} \right \}}$ and ${{\Big \langle}{\ldots}{\Big \rangle}_0}$ denotes expectation with respect to the Hamiltonian $\tilde{H}$ with $a=0$.

Noting $s_{i}={{\Big \langle}{z_i}{\Big \rangle}_a}={{\Big \langle}{z_i}{\Big \rangle}_0}$, from {{{Eq.}~{(\ref{firstrewrite})}}}, it follows
\begin{equation}
{G^{({1})}}={{\Big \langle}{H}{\Big \rangle}_0}=-\frac{1}{2}\sum_{i\neq j}c_{ij}^\ast s_{i} s_{j}^\ast.
\end{equation}
Then, from this equation and {{{Eq.}~{(\ref{firstrewrite})}}}, one has
\begin{equation}
{G{\left ( {a,{\left\{ {s_{i}} \right \}}} \right )}}={G^{({0})}}+a {{\Big \langle}{H}{\Big \rangle}_0}-\beta^{-1}\log{{\Big \langle}{\exp{\left\{ {a\beta B+\beta\sum_{n=2}\frac{a^n}{n!}A_{n}} \right \}}}{\Big \rangle}_0},{\label{secondrewrite}}
\end{equation}
where
\begin{equation}
B=\frac{1}{2}\sum_{i\neq j}c_{ij}^\ast{\left ( {z_i-s_{i}} \right )}{\left ( {z_j-s_{j}} \right )}^\ast.
\end{equation}

Evaluating ${G{\left ( {a,{\left\{ {s_{i}} \right \}}} \right )}}$ by expanding  this equation upto third order in $a$ yields
\begin{equation}
{G{\left ( {a,{\left\{ {s_{i}} \right \}}} \right )}}={G^{({0})}}+{{\Big \langle}{H}{\Big \rangle}_0}a-\frac{\beta}{2}{{\Big \langle}{B^2}{\Big \rangle}_0}a^2-\frac{\beta^2}{6}{{\Big \langle}{B^3}{\Big \rangle}_0}a^3+{\cal O}{\left ( {a^4} \right )},
\end{equation}
where it is noted that ${{\Big \langle}{B}{\Big \rangle}_0}={{\Big \langle}{A_{n}}{\Big \rangle}_0}={{\Big \langle}{B A_{n}}{\Big \rangle}_0}=0$ for every integer $n\ge 1$.
Then, noting ${{\Big \langle}{z_i-s_{i}}{\Big \rangle}_0}={{\Big \langle}{{\left ( {z_i-s_{i}} \right )}^\ast}{\Big \rangle}_0}=0$, we have
\begin{eqnarray}
&&{G{\left ( {a,{\left\{ {s_{i}} \right \}}} \right )}}\nonumber\\
&=&{G^{({0})}}+{\Bigg [}-\frac{1}{2}\sum_{i\neq j}c_{ij}^\ast s_{i} s_{j}^\ast{\Bigg ]} a\nonumber\\
&&+{\Bigg [}-\frac{\beta}{16}\sum_{i\neq j}{\Big \{}{E_{{i}}{\left ( {{2},{0}} \right )}}{E_{{j}}{\left ( {{0},{2}} \right )}}{c_{{ij}}^{\ast {2}}}+{E_{{i}}{\left ( {{0},{2}} \right )}}{E_{{j}}{\left ( {{2},{0}} \right )}}{c_{{ji}}^{\ast {2}}}\nonumber\\
&&+2{E_{{i}}{\left ( {{1},{1}} \right )}}{E_{{j}}{\left ( {{1},{1}} \right )}}{c_{{ij}}^{\ast {}}}{c_{{ji}}^{\ast {}}}{\Big \}}{\Bigg ]} a^2\nonumber\\
&&+{\Bigg [}-\frac{\beta^2}{96}\sum_{i\neq j}{\Big \{} {E_{{i}}{\left ( {{3},{0}} \right )}}{E_{{j}}{\left ( {{0},{3}} \right )}}{c_{{ij}}^{\ast {3}}}+{E_{{i}}{\left ( {{0},{3}} \right )}}{E_{{j}}{\left ( {{3},{0}} \right )}}{c_{{ji}}^{\ast {3}}} \nonumber\\
&&+3{E_{{i}}{\left ( {{2},{1}} \right )}}{E_{{j}}{\left ( {{1},{2}} \right )}}{c_{{ij}}^{\ast {2}}}{c_{{ji}}^{\ast {}}}+3{E_{{i}}{\left ( {{1},{2}} \right )}}{E_{{j}}{\left ( {{2},{1}} \right )}}{c_{{ij}}^{\ast {}}}{c_{{ji}}^{\ast {2}}}{\Big \}}\nonumber\\
&&-\frac{\beta^2}{48}\sum_{{\left ( {ijk} \right )}}{\Big \{}{E_{{i}}{\left ( {{2},{0}} \right )}}{E_{{j}}{\left ( {{1},{1}} \right )}}{E_{{k}}{\left ( {{0},{2}} \right )}}{c_{{ij}}^{\ast {}}}{c_{{ik}}^{\ast {}}}{c_{{jk}}^{\ast {}}}+{E_{{i}}{\left ( {{1},{1}} \right )}}{E_{{j}}{\left ( {{2},{0}} \right )}}{E_{{k}}{\left ( {{0},{2}} \right )}}{c_{{ik}}^{\ast {}}}{c_{{ji}}^{\ast {}}}{c_{{jk}}^{\ast {}}}\nonumber\\
&&+{E_{{i}}{\left ( {{1},{1}} \right )}}{E_{{j}}{\left ( {{1},{1}} \right )}}{E_{{k}}{\left ( {{1},{1}} \right )}}{c_{{ij}}^{\ast {}}}{c_{{jk}}^{\ast {}}}{c_{{ki}}^{\ast {}}}+{E_{{i}}{\left ( {{0},{2}} \right )}}{E_{{j}}{\left ( {{2},{0}} \right )}}{E_{{k}}{\left ( {{1},{1}} \right )}}{c_{{ji}}^{\ast {}}}{c_{{jk}}^{\ast {}}}{c_{{ki}}^{\ast {}}}\nonumber\\
&&+{E_{{i}}{\left ( {{2},{0}} \right )}}{E_{{j}}{\left ( {{0},{2}} \right )}}{E_{{k}}{\left ( {{1},{1}} \right )}}{c_{{ij}}^{\ast {}}}{c_{{ik}}^{\ast {}}}{c_{{kj}}^{\ast {}}}+{E_{{i}}{\left ( {{1},{1}} \right )}}{E_{{j}}{\left ( {{1},{1}} \right )}}{E_{{k}}{\left ( {{1},{1}} \right )}}{c_{{ik}}^{\ast {}}}{c_{{ji}}^{\ast {}}}{c_{{kj}}^{\ast {}}}\nonumber\\
&&+{E_{{i}}{\left ( {{1},{1}} \right )}}{E_{{j}}{\left ( {{0},{2}} \right )}}{E_{{k}}{\left ( {{2},{0}} \right )}}{c_{{ij}}^{\ast {}}}{c_{{ki}}^{\ast {}}}{c_{{kj}}^{\ast {}}}+{E_{{i}}{\left ( {{0},{2}} \right )}}{E_{{j}}{\left ( {{1},{1}} \right )}}{E_{{k}}{\left ( {{2},{0}} \right )}}{c_{{ji}}^{\ast {}}}{c_{{ki}}^{\ast {}}}{c_{{kj}}^{\ast {}}}{\Big \}}{\Bigg ]} a^3\nonumber\\
&&+{\cal O}{\left ( {a^4} \right )},{\label{complex}}
\end{eqnarray}
where ${E_{{i}}{\left ( {{n},{m}} \right )}}={{\Big \langle}{{\left ( {z_i-s_{i}} \right )}^n{\left ( {z_i^\ast-s_{i}^\ast} \right )}^m}{\Big \rangle}_0}$, and ${\left ( {ijk} \right )}$ denotes all combination to be taken so that either two of the indexes do not coincide (note that $(ij)$ implies $i\neq j$).
Then substituting {{{Eq.}~{(\ref{hebb})}}} into {{{Eq.}~{(\ref{complex})}}} yields, in the limit $N\rightarrow\infty$,
\begin{equation}
{G{\left ( {a,{\left\{ {s_{i}} \right \}}} \right )}}={G^{({0})}}-\frac{a}{2}\sum_{i\neq j}c_{ij}^\ast s_{i} s_{j}^\ast-\frac{\alpha N\beta{\left ( {1-q} \right )}^2}{8}a^2-\frac{\alpha N\beta^2{\left ( {1-q} \right )}^3}{24}a^3+{\cal O}{\left ( {a^4} \right )},
\end{equation}
where $q=\frac{1}{N}\sum_i {\left|{s_{i}}\right|}^2$.
Note that all the relevant terms higher than the term of first order in $a$ under  the limit $N\rightarrow\infty$ comes from the following in {{{Eq.}~{(\ref{complex})}}}
\begin{equation}
-\frac{a^n\beta^{n-1}}{2^n n}\sum_{{\left ( {i_1 i_2 \ldots i_n} \right )}}{E_{{i_1}}{\left ( {{1},{1}} \right )}}{E_{{i_2}}{\left ( {{1},{1}} \right )}}\ldots{E_{{i_n}}{\left ( {{1},{1}} \right )}}{c_{{i_1 i_2}}^{\ast {}}}{c_{{i_2 i_3}}^{\ast {}}}\ldots{c_{{i_n i_1}}^{\ast {}}}.{\label{remain}}
\end{equation}
Since every higher order term than the first order one contains $-a^n \beta^{n-1} {{\Big \langle}{B^n}{\Big \rangle}_0}/n!$, one may expect that it yields terms of the form of {{{Eq.}~{(\ref{remain})}}}.
Summarizing those terms we will have
\begin{eqnarray}
{G{\left ( {a,{\left\{ {s_{i}} \right \}}} \right )}}={G^{({0})}}-\frac{a}{2}\sum_{i\neq j}c_{ij}^\ast s_{i} s_{j}^\ast-\alpha N\sum_{n=2}\frac{\beta^{n-1}}{2^n n}{\left ( {1-q} \right )}^n a^n.
\end{eqnarray}

Then, noting $\partial q/\partial x_{i}+i\partial q/\partial y_{i}=2 s_{i}/N$,  we obtain
\begin{eqnarray}
&&R_i +i I_i\nonumber\\
&=&\frac{\partial G}{\partial x_{i}}+i\frac{\partial G}{\partial y_{i}}\nonumber\\
&=&\frac{\partial {G^{({0})}}}{\partial x_{i}}+i\ \frac{\partial {G^{({0})}}}{\partial y_{i}}-a\sum_{j\neq i} c_{ij} s_{j}-{\gamma^{{\mbox{\scriptsize TAP}}}} s_{i},{\label{aexpansion}}
\end{eqnarray}
where ${\gamma^{{\mbox{\scriptsize TAP}}}}=-a \alpha {\left\{ {a\beta{\left ( {1-q} \right )}/2} \right \}}/{\left\{ {1-a\beta{\left ( {1-q} \right )}/2} \right \}}$.

In the case of $a=0$ $\tilde{H}$ becomes
\begin{equation}
\tilde{H}=-\sum_i{\left\{ {{\left ( {\partial{G^{({0})}}/\partial x_i} \right )}\cos\phi_i+{\left ( {\partial{G^{({0})}}/\partial y_i} \right )}\sin\phi_i} \right \}}.
\end{equation}
Thus we have
\begin{eqnarray}
&&s_i\nonumber\\
&=&{{\Big \langle}{\cos\phi_i}{\Big \rangle}_0}+i{{\Big \langle}{\sin\phi_i}{\Big \rangle}_0}\nonumber\\
&=&\frac{I_1{\left ( {\beta\sqrt{{\left ( {\partial{G^{({0})}}/\partial x_i} \right )}^2+{\left ( {\partial{G^{({0})}}/\partial y_i} \right )}^2}} \right )}}{I_0{\left ( {\beta\sqrt{{\left ( {\partial{G^{({0})}}/\partial x_i} \right )}^2+{\left ( {\partial{G^{({0})}}/\partial y_i} \right )}^2}} \right )}}\nonumber\\
&&\times\frac{\partial{G^{({0})}}/\partial x_i+i\partial{G^{({0})}}/\partial y_i}{\sqrt{{\left ( {\partial{G^{({0})}}/\partial x_i} \right )}^2+{\left ( {\partial{G^{({0})}}/\partial y_i} \right )}^2}}\nonumber\\
&=&{f{\left ( {{0,\partial {G^{({0})}}/\partial x_{i}+i \partial {G^{({0})}}/\partial y_{i},{{{{\left ( {\partial {G^{({0})}}/\partial x_{i}+i \partial {G^{({0})}}/\partial y_{i}} \right )}}}^\ast}}} \right )}},{\label{azero}}
\end{eqnarray}
where $\displaystyle I_k(r)=\frac{1}{\sqrt{2\pi}}\int_0^{2\pi}\exp{\left ( {r\cos\varphi} \right )}\cos k\varphi \ d\varphi$, and ${f{\left ( {{0,h,h^\ast}} \right )}}$ is just the effective transfer function we introduced in {{{Eqs.}~{(\ref{circulate})}}},{{(\ref{density})}}, and {{(\ref{withd})}}.

Considering the case with $a=1$, from {{{Eqs.}~{(\ref{aexpansion})}}} and {{(\ref{azero})}} we finally obtain the TAP equation:
\begin{eqnarray}
s_{i}&=&{f{\left ( {{0,h_i^{{\mbox{\scriptsize TAP}}},h_i^{{\mbox{\scriptsize TAP}}\ast}}} \right )}},\\
h_i^{{\mbox{\scriptsize TAP}}}&=&\sum_{j\neq i} c_{ij} s_{j}+{\gamma^{{\mbox{\scriptsize TAP}}}} s_{i}+R_i+iI_i,\\
{\gamma^{{\mbox{\scriptsize TAP}}}}&=&-\alpha \frac{\beta{\left ( {1-q} \right )}/2}{1-\beta{\left ( {1-q} \right )}/2}.
\end{eqnarray}

\section{Derivation of {{{Eq.}~{(\ref{compare})}}}}{\label{evaluation}}

To derive {{{Eq.}~{(\ref{compare})}}}, we substitute  {{{Eq.}~{(\ref{plainlocalfield})}}} into {{{Eq.}~{(\ref{assumption})}}} to obtain
\begin{eqnarray}
&&h_i\nonumber\\
&=&m^1+\frac{1}{N}{\sum_{\mu> 1}\sum_{j\ne i}\xi_{i}^\mu{{{\xi_{j}^\mu}}^\ast}} s_j^\mu+{\gamma^{{\mbox{\scriptsize TAP}}}} s_i\nonumber\\
&&+\frac{\lambda}{N^2}\sum_{\mu>1}\sum_{j\ne i}\sum_{k\ne j}\xi^\mu_i{{{\xi_j^\mu}}^\ast}\xi_j^\mu{\left.\frac{\partial \tilde{f}}{\partial \tilde{h}}\right |_{{{\left ( {{\tilde{h_{{{j}}}^{{{\mu}}}},\tilde{h_{{{j}}}^{{{\mu}}}}^\ast}} \right )}}}}{{{\xi_k^\mu}}^\ast}s_k^\mu \nonumber\\
&&+\frac{{{{\lambda}}^\ast}}{N^2}\sum_{\mu>1}\sum_{j\ne i}\sum_{k\ne j}\xi^\mu_i{{{\xi_j^\mu}}^\ast}{{{\xi_j^\mu}}^\ast}{\left.\frac{\partial \tilde{f}}{\partial \tilde{{{{h}}^\ast}}}\right |_{{{\left ( {{\tilde{h_{{{j}}}^{{{\mu}}}},\tilde{h_{{{j}}}^{{{\mu}}}}^\ast}} \right )}}}}\xi_k^\mu {{{s_k^\mu}}^\ast}.{\label{substitution}}
\end{eqnarray}
Utilizing the relations $\xi_j^\mu{{{\xi_j^\mu}}^\ast}=1,\frac{1}{N}\sum_i\xi_i^\mu=0$, and so on, the fourth term of {{{Eq.}~{(\ref{substitution})}}} becomes, in the limit $N\rightarrow\infty$, 
\begin{eqnarray}
&&\frac{\lambda}{N^2}\sum_{\mu>1}\sum_{j\ne i}\sum_{k\ne j}\xi^\mu_i{{{\xi_j^\mu}}^\ast}\xi_j^\mu{\left.\frac{\partial \tilde{f}}{\partial \tilde{h}}\right |_{{{\left ( {{\tilde{h_{{{j}}}^{{{\mu}}}},\tilde{h_{{{j}}}^{{{\mu}}}}^\ast}} \right )}}}}{{{\xi_k^\mu}}^\ast}s_k^\mu \nonumber\\
&=&\frac{\lambda}{N}\sum_{\mu>1}\sum_k\xi_i^\mu{{{\xi_k^\mu}}^\ast}s_k^\mu\frac{1}{N}\sum_{j\ne i}{\left.\frac{\partial \tilde{f}}{\partial \tilde{h}}\right |_{{{\left ( {{\tilde{h_{{{j}}}^{{{\mu}}}},\tilde{h_{{{j}}}^{{{\mu}}}}^\ast}} \right )}}}}-\frac{\lambda}{N}\sum_{\mu>1}\xi_i^\mu\frac{1}{N}\sum_{j\ne i}{\left.\frac{\partial \tilde{f}}{\partial \tilde{h}}\right |_{{{\left ( {{\tilde{h_{{{j}}}^{{{\mu}}}},\tilde{h_{{{j}}}^{{{\mu}}}}^\ast}} \right )}}}}{{{\xi_j^\mu}}^\ast} s_j^\mu\nonumber\\
&=&\frac{u\lambda}{N}\sum_{\mu>1}\sum_{k\ne i}\xi_i^\mu{{{\xi_k^\mu}}^\ast}s_k^\mu+\alpha u\lambda s_i.
\end{eqnarray}
Following the almost same scheme the fifth term of {the right hand side}\ of {{{Eq.}~{(\ref{substitution})}}} is shown to  vanish in the limit $N\rightarrow\infty$.
Substituting {{{Eq.}~{(\ref{assumption})}}} into {the left hand side}\ of {{{Eq.}~{(\ref{substitution})}}} we obtain {{{Eq.}~{(\ref{compare})}}}.

\section{Derivation of {{{Eqs.}~{(\ref{scu})}}} and {{(\ref{scr})}}}{\label{miscellaneous}}

The {{{Eq.}~{(\ref{scu})}}} is straightforwardly derived from the definition of $u$ by noting ${{\Bigg \langle}{\Bigg \langle}{\frac{\partial \tilde{f}}{\partial \tilde{h}}}{\Bigg \rangle}{\Bigg \rangle}}={{\Bigg \langle}{\Bigg \langle}{\frac{1}{2}{\left [ {\frac{\partial\tilde{f}}{\partial{\left\{ {{\mbox{Re}}{\left ( { \tilde{h}} \right )}} \right \}}}-i\frac{\partial\tilde{f}}{\partial{\left\{ {{\mbox{Im}}{\left ( {\tilde{h}} \right )}} \right \}}}} \right ]}}{\Bigg \rangle}{\Bigg \rangle}}$ and performing integration by parts.
To show {{{Eq.}~{(\ref{scr})}}} from {{{Eq.}~{(\ref{r})}}} it is suffice to prove that $u$ is real.

To show $u$ is real, note the rotationary symmetry structure of the form of transfer function~{{(\ref{circulate})}} as is illustrated in {{{Fig.}~\ref{field}}}.
Because of this symmetry structure of ${f{\left ( {{\omega,\overline{h},{{{\overline{h}}}^\ast}}} \right )}}$ we also have ${\tilde{f}{\left ( {{\omega,r e^{i\theta},{{{{\left ( {r e^{i\theta}} \right )}}}^\ast}}} \right )}}=e^{i\theta}{\tilde{f}{\left ( {{\omega,r,r}} \right )}}$ in the presence of non-zero complex ${\gamma^{{\mbox{\scriptsize TOTAL}}}}$.
One also immediately finds $f{\left ( {\omega,r,r} \right )}={{{f{\left ( {-\omega,r,r} \right )}}}^\ast}$ and ${\tilde{f}{\left ( {{\omega,r,r}} \right )}}={{{{\tilde{f}{\left ( {{-\omega,r,r}} \right )}}}}^\ast}$.
Then it follows that ${\tilde{f}{\left ( {{\omega,r e^{i\theta},{{{{\left ( {r e^{i\theta}} \right )}}}^\ast}}} \right )}}={{{{\tilde{f}{\left ( {{-\omega,{{{{\left ( {r e^{i\theta}} \right )}}}^\ast},r e^{i\theta}}} \right )}}}}^\ast}$ and ${\left ( {x-i y} \right )}{\tilde{f}{\left ( {{\omega,\tilde{h},{{{\tilde{h}}}^\ast}}} \right )}}={{{{\left\{ {{\left ( {x+i y} \right )}{\tilde{f}{\left ( {{-\omega,{{{\tilde{h}}}^\ast},\tilde{h}}} \right )}}} \right \}}}}^\ast}$.
On the other hand noting $g{\left ( {\omega} \right )}=g{\left ( {-\omega} \right )}$ and changing the variables for integration, we have, from {{{Eq.}~{(\ref{scu})}}},
\begin{equation}
\sqrt{\alpha r}u={{\Bigg \langle}{\Bigg \langle}{{\left ( {x+i y} \right )}{\tilde{f}{\left ( {{-\omega,\tilde{h}^\ast,\tilde{h}}} \right )}}}{\Bigg \rangle}{\Bigg \rangle}}.
\end{equation}
Accordingly we have
\begin{equation}
\sqrt{\alpha r}u={{\Bigg \langle}{\Bigg \langle}{{\left\{ {{\left ( {x-i y} \right )}{\tilde{f}{\left ( {{\omega,\tilde{h},\tilde{h}^\ast}} \right )}}} \right \}}^\ast}{\Bigg \rangle}{\Bigg \rangle}}={{\sqrt{\alpha r}u}^\ast}
\end{equation}
to conclude that $u$ is real.

\section{Derivation of the macroscopic order parameter equations for the case with $D=0$ and $\alpha=0$}{\label{a0}}

In the case with $D=0$ and $\alpha=0$, substituting {{{Eq.}~{(\ref{nod})}}} into {{{Eq.}~{(\ref{finite})}}}, we have 
\begin{equation}
m=\left\{\begin{array}{lc}
	a&0<m\le\omega_1\\
	a+(1-a)\frac{\sqrt{m^2-\omega_1^2}}{m}& \omega_1<m
\end{array}\right . .{\label{a0scm}}
\end{equation}
Using {{{Eq.}~{(\ref{nod})}}} we also obtain, from {{{Eqs.}~{(\ref{scq})}}} and {{(\ref{scr})}}, $q$ and $u$ as a function of $m$:
\begin{eqnarray}
q&=&\left\{\begin{array}{lc}
	a+(1-a){\left ( {\frac{\omega_1-\sqrt{\omega_1^2-m^2}}{m}} \right )}^2&0<m\le\omega_1\\
	1& \omega_1<m
\end{array}\right. ,{\label{a0scq}}\\
u&=&\left\{\begin{array}{lc}
	\frac{a}{2m}&0<m\le\omega_1\\
	\frac{a}{2m}+\frac{1-a}{2\sqrt{m^2-\omega_1^2}}& \omega_1<m
\end{array}\right. ,{\label{a0scu}}
\end{eqnarray}
where we have noted $u={{\Bigg \langle}{\Bigg \langle}{{\mbox{Re}}{\left\{ {\frac{\partial f}{\partial h}} \right \}}}{\Bigg \rangle}{\Bigg \rangle}}={{\Bigg \langle}{\Bigg \langle}{{\mbox{Re}}{\left\{ {\frac{e^{-i\theta}}{2}{\left ( {\frac{\partial f}{\partial r}-\frac{i}{r}\frac{\partial f}{\partial \theta}} \right )}} \right \}}}{\Bigg \rangle}{\Bigg \rangle}}$, that is obtained by representing the local field with the polar coordinate. {i e. }\  $h=r e^{i\theta}$.

\begin{figure}
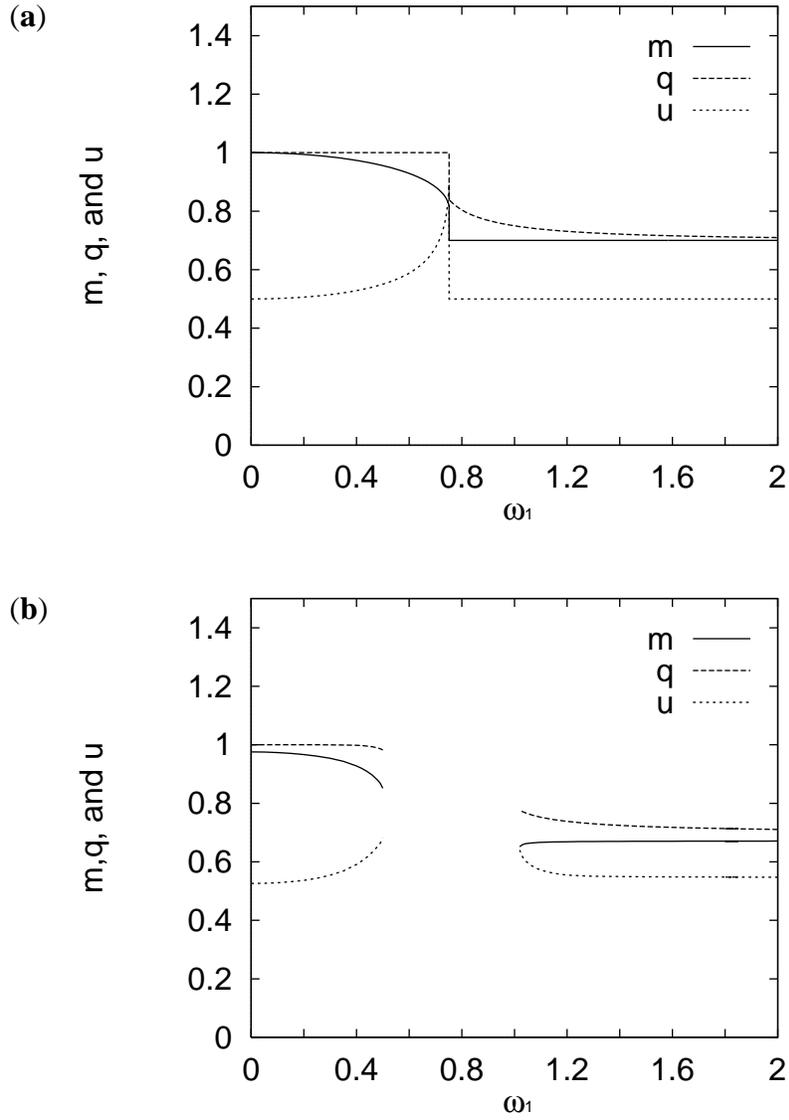

{}{}
{}{}
{}
\caption{{{\bf (a)}}\ $\omega_1$-dependences of the order parameters $m,q,$ and $u$ obtained from {{{Eqs.}~{(\ref{a0scm})}}},{{(\ref{a0scq})}}, and {{(\ref{a0scu})}} are displayed in the case with $a=0.7$ and $\alpha=0$.
{{\bf (b)}}\ same as {{\bf (a)}}\ for the case with $\alpha=0.02$.
The gap separating two types of retrieval state (the large $\omega_1$ regimes and small $\omega_1$ regimes) implies the disappearance of retrieval states.}
{\label{reason}}
\end{figure}

As $\omega_1$ approaches the point of phase transition from below, $u$ increases as is shown in {{{Fig.}~\ref{reason}}}{{\bf (a)}} .
At the phase transition point where $m=a+{\left ( {1-a} \right )}\sqrt{m^2-\omega_1^2}/m$ and $\frac{\partial}{\partial m}{\left\{ {a+{\left ( {1-a} \right )}\sqrt{m^2-\omega_1^2}/m} \right \}}=1$,  one has $u=1$.

\vspace*{\fill}{\noindent\texttt{\footnotesize Compiled on \today.}}

\label{lastpage}

{\newpage}

\begin{figure}
{}{}
\caption{Graphical representation of the effective transfer function ${f{\left ( {{\omega,\overline{h},\overline{h}^\ast}} \right )}}$ in the case with $D=0$ and $\omega>0$.
The output ${f{\left ( {{\omega,\overline{h},\overline{h}^\ast}} \right )}}$ is represented  by a vector at the position $\overline{h}$ on the complex plane.
In the region, where ${\left|{\overline{h}}\right|}<1$, oscillators get synchronized with ${\left|{{f{\left ( {{\omega,\overline{h},\overline{h}^\ast}} \right )}}}\right|}=1$, while inside the circle oscillators get desynchronized with ${\left|{{f{\left ( {{\omega,\overline{h},\overline{h}^\ast}} \right )}}}\right|}<1$.
In the case of $\omega<0$ the rotational direction of the flow pattern gets reversed owing to the property ${f{\left ( {{-\omega,{\overline{h}},{\overline{h}}^\ast}} \right )}}={\left\{ {{f{\left ( {{\omega,{\overline{h}}^\ast,{\overline{h}}}} \right )}}} \right \}}^\ast$.}
{\label{field}}
\end{figure}

\begin{thebibliography}{10}

\bibitem{gray}
C.~M. Gray and W.~Singer.
\newblock {\em Proc. Natl. Acad. Sci. USA}, 86:1698, 1989.

\bibitem{eckhorn}
R.~Eckhorn, R.~Bauer, W.~Jordan, M.~Brosch, W.~Kruse, M.~Munk, and R.~J.
  Reitboeck.
\newblock {\em Biol. Cybern.}, 60:121, 1988.

\bibitem{malsburg}
C.~von~der Malsburg and W.~Schneider.
\newblock {\em Biol. Cybern.}, 54:29, 1986.

\bibitem{malsburg2}
C.~von~der Malsburg and Ber. Bunsenges.
\newblock {\em Phys. Chem}, 89:703, 1985.

\bibitem{milner}
P.~M. Milner.
\newblock {\em Psycholog. Rev.}, 81:521, 1974.

\bibitem{damasio}
A.~R. Damasio and Semin.
\newblock {\em Neurosci}, 2:287, 1990.

\bibitem{hopfield}
J.~J. Hopfield.
\newblock {\em Proc. Natl. Acad. Sci. USA}, 79:2554, 1982.

\bibitem{amit}
D.~J. Amit, H.~Gutfreund, and H.~Sompolinsky.
\newblock {\em Ann. Phys.}, 173:30, 1987.

\bibitem{shiino}
M.~Shiino and T.~Fukai.
\newblock {\em J. Phys. A: Math. Gen.}, 23:L1009, 1990.

\bibitem{kuhn}
R.~Kuhn, S.~Bos, and J.~L. {van Hemmen}.
\newblock {\em Phys. Rev. A}, 43:2084, 1991.

\bibitem{kuhn2}
{\em J. Phys. A}, 22, 1989.

\bibitem{gardner}
E.~Gardner.
\newblock {\em J. Phys. A: Math. Gen.}, 21:257, 1988.

\bibitem{kuhn3}
R.~Kuhn and S.~Bos.
\newblock {\em J. Phys. A}, 26:831, 1993.

\bibitem{coolen}
A.~C.~C. Coolen and D.~Sherrington.
\newblock {\em Phys. Rev. E}, 49:1921, 1994.

\bibitem{fukai}
T.~Fukai and M.~Shiino.
\newblock {\em Phys. Rev. A}, 42:7459, 1990.

\bibitem{waugh}
F.~R. Waugh and C.~M. Marcus.
\newblock {\em Phys. Rev. Lett.}, 64:1986, 1990.

\bibitem{cugliandolo}
L.~F. Cugliandolo and M.~V. Tsodkys.
\newblock {\em J. Phys. A: Math. Gen.}, 27:741, 1994.

\bibitem{cugliandolo2}
L.~F. Cugliandolo.
\newblock {\em Neural Comp.}, 6:220, 1994.

\bibitem{griniasty}
M.~Griniasty, M.~V. Tsodks, and D.~J. Amit.
\newblock {\em Neural Comp.}, 5:1, 1993.

\bibitem{peretto}
P.~Peretto.
\newblock {\em J. Phys. France}, 49:711, 1988.

\bibitem{hebb}
D.~O. Hebb.
\newblock {\em Organization of Behavior}.
\newblock Wiley, NewYork, 1949.

\bibitem{yoshioka}
M.~Yoshioka and M.~Shiino.
\newblock {\em Phys. Rev. E}, 55:7401, 1997.

\bibitem{yoshioka2}
M.~Yoshioka and M.~Shiino.
\newblock {\em J. Phys. Soc. Jpn.}, 66:1294, 1997.

\bibitem{mass}
W.~Mass.
\newblock {\em Neural Comp.}, 8:1, 1996.

\bibitem{hansel}
D.~Hansel, G.~Mato, and C.~Meunier.
\newblock {\em Neural Comp.}, 7:307--337, 1995.

\bibitem{vreeswijk}
C.~van Vreeswijk, L.~F. Abbott, and G.~B. Ermentrout.
\newblock {\em J. Comp. Neurosci.}, 1:313--321, 1994.

\bibitem{gerstner}
W.~Gerstner, R.~Ritz, and J.~L. {van Hemmen}.
\newblock {\em Bio. Cybern.}, 69:503, 1993.

\bibitem{gerstner2}
W.~Gerstner.
\newblock {\em Phys. Rev. E}, 51:738--758, 1995.

\bibitem{herz}
A.~V.~M. Herz, Z.~Li, and J.~L. {van Hemmen}.
\newblock {\em Phys. Rev. Lett.}, 66:1370, 1991.

\bibitem{treves}
A.~Treves, E.~T. Rolls, and M.~W. Simmen.
\newblock {\em Physica D}, 107:392, 1997.

\bibitem{bressloff1}
P.~C. Bressloff and S.~Coombes.
\newblock {\em Phys. Rev. Lett.}, 81:2168, 1998.

\bibitem{bressloff2}
P.~C. Bressloff and S.~Coombes.
\newblock {\em Phys. Rev. Lett.}, 81:2384, 1998.

\bibitem{carson}
C.~C. Chow.
\newblock {\em Physica D}, 118:343, 1998.

\bibitem{hodgkin}
A.~L. Hodgkin and A.~F. Huxley.
\newblock {\em J. Physiol.}, 117:500, 1952.

\bibitem{fitzhugh}
R.~FitzHugh.
\newblock {\em Biophys. J.}, 1:445.

\bibitem{nagumo}
J.~Nagumo, S.~Arimoto, and S.~Yoshizawa.
\newblock {\em Proc. IRE}, 50:2061.

\bibitem{yoshioka3}
M.~Yoshioka and M.~Shiino.
\newblock {\em Phys. Rev. E}, 58:3, 1998.

\bibitem{kuramoto}
Y.~Kuramoto.
\newblock {\em Chemical oscillations, waves, and turbulence}.
\newblock Springer-Verlag, 1984.

\bibitem{sakaguchi}
H.~Sakaguchi.
\newblock {\em Prog. Theor. Phys.}, 79:39, 1988.

\bibitem{daido}
H.~daido.
\newblock {\em J. Stat. Phys.}, 60:753, 1990.

\bibitem{arenas}
A.~Arenas and C.~J. {Perez Vicente}.
\newblock {\em Europhys. Lett.}, 26(2):79, 1994.

\bibitem{cook}
J.~Cook.
\newblock {\em J. Phys. A: Math. Gen.}, 22:2057, 1989.

\bibitem{aoyagi}
T.~Aoyagi and K.~Kitano.
\newblock {\em Phys. Rev. E}, 55:7424, 1997.

\bibitem{yamana}
M.~Yamana, M.~Shiino, and M.~Yoshioka.
\newblock {\em cond-mat/9901301}.

\bibitem{shiino2}
M.~Shiino and T.~Fukai.
\newblock {\em J. Phys. A: Math. Gen.}, 25:L375, 1992.

\bibitem{shiino3}
M.~Shiino and T.~Fukai.
\newblock {\em Phys. Rev. E}, 48:867, 1993.

\bibitem{thouless}
D.~J. Thouless, P.~W. Anderson, and R.~G. Palmer.
\newblock {\em Phil. Mag.}, 35:593, 1977.

\bibitem{mezard}
M.~Mezard, G.~Parisi, and M.~A. Virasoro.
\newblock {\em Spin glass theory and beyond}.
\newblock World Scientific, 1987.

\bibitem{kirkpatrick}
S.~Kirkpatrick and D.~Sherrington.
\newblock {\em Phys. Rev. B}, 17:4384, 1978.

\bibitem{plefka}
T.~Plefka.
\newblock {\em J. Phys. A: Math. Gen.}, 15:1971, 1982.

\bibitem{nakanishi}
K.~Nakanishi and H.~Takayama.
\newblock {\em J. Phys. A: Math. Gen.}, 30:1997, 8085.

\bibitem{park}
K.~Park and M.~Y. Choi.
\newblock {\em Phys. Rev. E}, 52:2907, 1995.

\bibitem{aonishi2}
T.~Aonishi, K.~Kurata, and M.~Okada.
\newblock {\em Phys. Rev. Lett.}, 82:2800--2803, 1999.

\bibitem{xiao}
X.~J. Wang and G.~Buzaki.
\newblock {\em J. Neurosci.}, 16:6402, 1996.

\bibitem{terman}
D.~Terman, N.~Kopell, and A.~Bose.
\newblock {\em Physica D}, 117:241, 1998.

\bibitem{gabbay}
M.~Gabbay, E.~Ott, and P.~N. Guzdar.
\newblock {\em Physica D}, 118:371, 1998.

\bibitem{ernst}
U.~Ernst, K.~Pawelzik, and T.~Geisel.
\newblock {\em Phys. Rev. Lett}, 74:1570, 1995.

\end{thebibliography}
\end{document}